\documentclass[twocolumn]{revtex4}
\usepackage{amsmath,amssymb}
\usepackage{graphicx}
\usepackage{times,txfonts}
\usepackage{color}
\usepackage{natbib}

\newcommand{\abs}[1]{\left| #1 \right|}

\newcommand{\ket}[1]{\left| #1 \right\rangle}
\newcommand{\braket}[2]{\left\langle {#1{\left| \vphantom{#1 #2} \right.} #2} \right\rangle}
\renewcommand{\epsilon}{\varepsilon}
\renewcommand{\phi}{\varphi}

\begin{document}
\begin{titlepage}
	\title{Towards Communication in a Curved Spacetime Geometry}
	\author{Qasem Exirifard$^{*}$}
	\author{Eric Culf}
	\author{Ebrahim Karimi$^{*}$}
	\affiliation{Department of Physics, University of Ottawa, 25 Templeton St., Ottawa, Ontario, K1N 6N5 Canada}
	\affiliation{$^{*}$Corresponding authors: qexirifa@uottawa.ca and ekarimi@uottawa.ca}
	\begin{abstract}
		 The current race in quantum communication -- endeavouring to establish a global quantum network -- must account for special and general relativistic effects. The well-studied general relativistic effects include Shapiro time-delay, gravitational lensing, and frame dragging which all are due to how a mass distribution alters geodesics. Here, we report how the curvature of spacetime geometry affects the propagation of information carriers along an arbitrary geodesic. An explicit expression for the distortion onto the carrier wavefunction in terms of the Riemann curvature is obtained. Furthermore, we investigate this distortion for  anti de Sitter and  Schwarzschild geometries. For instance, the spacetime curvature causes a 0.10~radian phase-shift for communication between Earth and the International Space Station on a monochromatic laser beam and quadrupole astigmatism; can cause a  12.2 \% cross-talk between structured modes traversing through the solar system. Our finding shows that this gravitational distortion is significant, and it needs to be either pre- or post-corrected at the sender or receiver to retrieve the information.
	\end{abstract}
	
	\maketitle
	
\end{titlepage}

\section*{Introduction} Photons, electromagnetic waves, are widely used in classical and quantum communication since they do not possess electric charge or rest mass. However, a photon's traits, e.g., group and phase velocity, wavelength, linear and optical angular momentum, are modified inside or during propagation through a linear or a nonlinear medium. These traits are governed by Maxwell's equations which are the relativistic quantum field theory of the $U(1)$ gauge connection. Understanding how these optical properties are altered upon propagation is a key element for any optical communication network. In optical communication, the sender and the receiver, namely Alice and Bob, use one or several internal photonic degrees of freedom, such as wavelength, polarisation, transverse mode, or time-bins, to share information, including a ciphertext and the secret key to decrypt the ciphertext. The propagation, e.g., through fibre, air or underwater channels, causes these photonic degrees of freedom to be altered, and thus causes undesired errors on the shared information. Therefore, the alteration to those degrees of freedom for any communication channel needs to be considered and well-examined. Sharing information with a longer range or with moving objects, e.g. satellites, airplanes, submersibles~\cite{Ursin:07,Yin:17,Yin:17PRL,QKD1, QKD2}, requires the optical beam to not only traverse through a medium, but in a few cases, also in the fabric of the spacetime geometry, where general relativistic effects manifest~\cite{Einstein:11,Pound:60,Chou:10,GPS}. Effects associated to the change of the geodesic due to a mass distribution, such as Shapiro time-delay~\cite{Shapiro:68}, gravitational lensing~\cite{Lensing}, and frame dragging~\cite{Everitt:2011hp}, are well-studied and observed \cite{GRtest,Liu:2019rer,Kish:2018cxk,Pierini:2018naq}. 

We explore the propagation of relativistic wavepackets along an arbitrary null geodesic in a general curved spacetime geometry, and show how the curvature of the spacetime geometry distorts the wavepacket as it travels along the null geodesic. Different methods are used to tackle this study. For instance, the four-dimensional Klein-Gordon equation is approximated to a simple two-dimensional partial differential equation by ignoring all the multi-polar modes ~\cite{Bruschi:2013sua, Bruschi:2014cma}. In particular, in derivation of eq.(7) from eq.(6) in \cite{Bruschi:2013sua},~the second term on the left-hand side of eq. (41) of ~\cite{Jonsson:2020npo} has not been taken into account, therefore, \cite{Bruschi:2013sua, Bruschi:2014cma} cannot claim to reproduce all the effects of a curved space-time geometry. In \cite{Jonsson:2020npo}, all the multi-polar $\ell$ modes are presented only at the level of the equations, however, the upper value of $\ell=100$ on the multi-polar modes is considered to compute the solution. On the surface of the earth, a narrow beam with an initial width of $10~cm$ and a large value of Rayleigh range requires taking into account the contribution of multi-polar modes up to at-least $\ell=10^9$. So the solutions presented in ref. \cite{Jonsson:2020npo} do not take into consideration all the multi-polar modes required to calculate the effects of the curvature at the vicinity of the Earth. Here, we present a computationally simple method to calculate the distortion by the curvature of any spacetime geometry on any localised wavepacket.

In flat spacetime geometry, the propagation of each polarisation of photon is isomorphic to the propagation of a massless scalar field. Based on our understanding of the Einstein equivalence principle, we expect that studying a massless scalar field theory also captures some features of a photon’s propagation in curved spacetime geometry. Therefore, in this study, we consider the propagation of a massless scalar in the bulk of the paper. In Supplementary Note 2, we prove that in the Lorentz gauge, each linear polarisation of the photon in a curved spacetime geometry gets corrected as if it were a massless scalar field. It is reported that the Riemann tensor quantum mechanically alters the wavepacket propagating along a geodesic. The alteration operators depend on the geodesic and components of the Riemann tensor on the geodesic. The alteration is calculated for examples including the space-time geometry around the Earth and the Sun.

\begin{figure*}[t]
	\begin{center}
		\includegraphics[width=2\columnwidth]{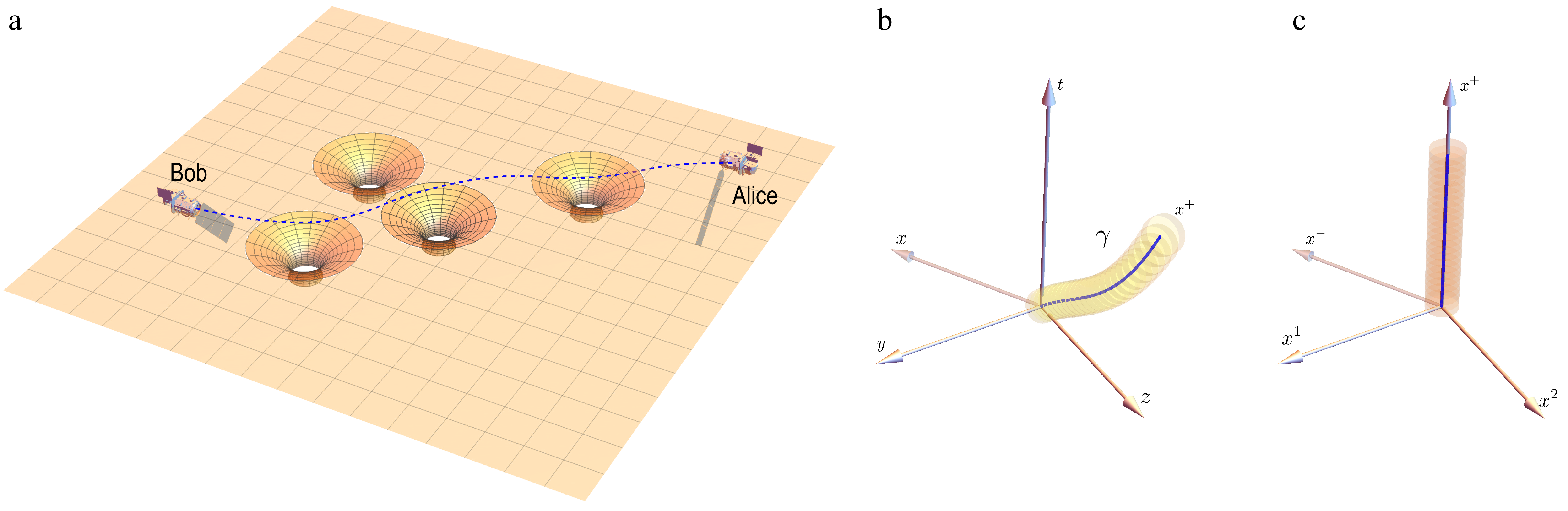}
		\caption[]{{\bf Schematics of communication in a general curved spacetime geometry and proper chosen coordinates.} {\bf a} Two parties, namely Alice and Bob, communicate in a general curved spacetime geometry. Alice encodes her message in a sequence of information carriers  and sends them to Bob. The information traverses through the spacetime over a geodesic $\gamma$. The physical traits of information are distorted by the curvature of the space-time, causing errors to the field wavepacket. {\bf b} The shown null geodesic  $\gamma$. Close to $\gamma$, at the local coordinates, the metric is approximately pseudo-Euclidean. {\bf c} The associated Fermi coordinates, the null-geodesic-path is mapped to $x^{+}$.}
		\label{fig:fig1}
	\end{center}
\end{figure*}
\section*{Results and Discussion}  We start by considering a  relativistic massless scalar field $\psi:=\psi(x^\mu)$ that propagates in a curved spacetime geometry $x^\mu=(t,x^i)=(t,x,y,z)$ with an arbitrary Riemann curvature tensor $R_{abcd}$. The units are chosen such that the speed of light in vacuum and the reduced Planck constant are equal to one, i.e. $c=1$ and $\hbar=1$.  Let us consider a localized wave function (information carrier) whose size is small compared to the curvature of the spacetime geometry. At the leading order, therefore, the carrier can be treated as a massless point-like particle that travels along a null geodesic $\gamma$, see Fig.~\ref{fig:fig1} {\bf a}.   We choose the local Fermi coordinates~\cite{Fermi:22} along the geodesics in order to compute the quantum relativistic corrections. The metric's components in the Fermi coordinates can be expanded in terms of the components of the Riemann tensor $R_{abcd}$ and its covariant derivatives evaluated on the geodesic, see Fig.~\ref{fig:fig1} {\bf b},{\bf c}. The expansion of the metric up to quadratic order in the transverse coordinates of the geodesic of a massless particle $\gamma$ is given by~\cite{Blau:06},
\begin{eqnarray}\label{eq1}
	ds^2 &=& 2 dx^+ dx^- + \delta_{ab} dx^a dx^b- \bigg[ R_{+\bar{a}+\bar{b}} x^{\bar{a}} x^{\bar{b}} (dx^+)^2 \\ \nonumber
	&+&  \frac{4}{3}R_{+\bar{b}\bar{a}\bar{c}} x^{\bar{b}}x^{\bar{c}} (dx^+ dx^{\bar{a}})+\frac{1}{3} R_{\bar{a}\bar{c}\bar{b}\bar{d}}
	x^{\bar{b}}x^{\bar{c}} (dx^{\bar{a}} dx^{\bar{b}})\bigg]+\ldots,
\end{eqnarray}
where $x^{\pm}=(x^3\pm t)/{\sqrt{2}}$ represent the Dirac light-cone coordinates~\cite{Dirac:49} in the Fermi coordinates ($x^+$ is always tangent to the null geodesic Fig. \ref{fig:fig1}-{\bf b}), $\delta_{ab}$ is the Kronecker delta, $\hat{\dot{\gamma}}$ is the tangent of the null geodesic and $a,b \in \{1,2\}$, and $(x^{\bar{a}}) = (x^-, x^{a})$ and the curvature components are evaluated on $\gamma$.  {The tree-level action of a massless scalar field $\psi$ in a general curved spacetime geometry is given by $S[\psi]= \frac{1}{2} \int dx^4 \sqrt{-g} g^{\mu\nu} \partial_\mu \psi \partial_\nu \psi$, where $g$ is the determinant of the metric $g_{\mu\nu}$, and $\mu,\nu\in\{\pm,a\}$. Since the tree-level action is quadratic in terms of $\psi$, it is quantum mechanically exact, which can be verified by looking at its generating function, i.e., $Z[J]= {\int {\cal D}\psi e^{-i(S[\psi]+\int d^4x \sqrt{-\det g}J \psi)}}/{\int {\cal D}\psi e^{-i S[\psi]}}$ --  ${\cal D}\psi$ represents the integration over all field configurations and $J$ is the source field. Its exact effective action, as defined by the Legendre transformation of $\ln Z$ , coincides with the tree-level action, i.e., $\Gamma[\psi_c]= S[\psi_c]$. For a general action, $\phi_c$ resembles a ``classical'' field whose action is given by $\Gamma[\psi_c]$, while $\Gamma[\psi_c]$ encapsulates all the quantum loop corrections}. The exact effective action includes both the classical and quantum effects. The classical effects are those that can be reproduced by motion of a point-like particle along the geodesic; the rest are quantum. The effective action of a free photon propagating in curved spacetime geometry coincides to the tree-level action, therefore, we omit the subscript $c$.  

The massless scalar {(quantum)} field $\psi$ obeys the (covariant) wave equation, $\Box\psi=(-g)^{-\frac{1}{2}}\partial_\mu (-g)^{\frac{1}{2}}g^{\mu\nu} \partial_\nu\psi=0$. In the Fermi coordinates, $g_{\mu\nu}$ can be viewed as a perturbation to the Minkowski metric, inducing expansion series for the inverse and determinant of the metric: $g^{\mu\nu}=\eta^{\mu\nu}+\epsilon\delta\!g^{\mu\nu}+O(\epsilon^2)$ and $\ln(\sqrt{-\det g})=\epsilon \delta\!g + O(\epsilon^2)$. $\epsilon$ is the systematic perturbation parameter introduced to keep track of the perturbation series, which means that all the components of the Riemann tensor in Eq.~\eqref{eq1} are multiplied with $\epsilon$, and $\epsilon$ is treated as an infinitesimal parameter. At the end of the computation, we set $\epsilon=1$. This technique helps us to systematically perform perturbations for small curvatures. Utilizing the perturbation  gives,
\begin{equation}\label{Eq2}
	\Box \psi = \Box^{(0)} \psi  + \epsilon ( \partial_\mu (\delta g^{\mu\nu} \partial_\nu \psi) + \eta^{\mu\nu} \partial_\nu \delta g \partial_\mu \psi ) = O(\epsilon^2),
\end{equation}
where $\Box^{(0)}=\eta^{\mu\nu} \partial_\mu \partial_\nu = 2\partial_-\partial_++\nabla^2_{\!\!\perp}$ is the d'Alembert operator in the flat spacetime geometry, and $\nabla^2_{\!\!\perp}=\partial_1^2+\partial_2^2$. The perturbative nature of Eq.~\eqref{Eq2} seeks for a series expansion, $\psi=\psi^{(0)}+\epsilon \psi^{(1)} + O(\epsilon^2)$. Here, $\psi^{(0)}$ satisfies the scalar wave equation in the flat spacetime geometry $\Box^{(0)}\psi^{(0)} = 0$, and the perturbed term to the wavefunction, $\psi^{(1)}$, yields,
\begin{equation}\label{Eq3}
	\Box^{(0)}\psi^{(1)}=-\partial_\mu(\delta\!g^{\mu\nu}\partial_\nu\psi^{(0)})+\eta^{\mu\nu} \partial_\nu \delta\!g \partial_\mu\psi^{(0)} \,.
\end{equation}
We assume the Fourier expansion in terms of $x^-$ variable, $\psi^{(0)}=\int\!d\omega f^{(0)}_\omega(x^+,x^a) e^{i \omega x^-}$, where $f^{(0)}_\omega$ satisfies the paraxial equation $\left(2i \omega \partial_+ +\nabla^2_{\!\!\perp}\right)f_\omega^{(0)} = 0$. This implies that the solutions, given by the paraxial approximation in optics~\cite{siegman:86,Iwo:06}, are exact.  The paraxial equation is isomorphic to the Schr\"odinger equation, and its solutions (the transverse and longitudinal parts) can be expressed in the form of Laguerre-Gauss (LG) modes (with an azimuthally-symmetric intensity profile) or Hermite-Gaussian (HG) wavepackets~\cite{Bliokh:07}. We consider a wavepacket wherein the field is slowly varying, and assume that the metric does not significantly change inside the wavepacket. Therefore, all derivatives of $\partial_\mu \psi^{(0)}$, except $\partial_- \psi^{(0)}$, are negligible, and the leading term in the right hand side of Eq.~\eqref{Eq3} is $\partial_- (\delta\!g^{--}\partial_- \psi)$, where $\delta\!g^{--}= - g_{++}^{(1)}= R_{+\bar{a}+\bar{b}} x^{\bar{a}} x^{\bar{b}}$. Therefore, Eq.~\eqref{Eq3} reduces to,
\begin{equation}
	\label{Eq4}
	\Box^{(0)} \psi^{(1)}=-R_{+\bar{a}+\bar{b}}~x^{\bar{a}} x^{\bar{b}}\partial_{-}^2\psi^{(0)}.
\end{equation}
Here, $\psi^{(1)} =\int d\omega  f^{(1)}_\omega(x^+,x^a) e^{i\omega x^-}$ with $f^{(1)}_\omega$ being the correction to the structure function for the frequency of $\omega$. We have found the solutions to Eq.~\eqref{Eq4} -- see the Supplementary Note 1 for more detail on the derivation. The solution is, \\ 
\begin{eqnarray*}
	\psi(x^\mu)=\int\!\!d\omega\!\sum_{p,\ell,n}\!c_{p\ell n}(\omega) e^{i\omega x^-} \left(1+\epsilon ({ \cal O}^\omega+ {\cal Q}_U^\omega + {\cal Q}_N^\omega) \right)f^{(0)}_\omega,
\end{eqnarray*}
where $c_{p,\ell,n}(\omega)$ are defined based on the initial and boundary conditions. The operators, ${ \cal O}^\omega$, ${\cal Q}_U^\omega$ and ${\cal Q}_N^\omega$ encodes how the curvature of the space-time geometry distorts the wavepacket. They are given by,
\begin{eqnarray}\label{Eq5}
	{\cal O}^\omega &=& - \frac{i\omega}{2}  {\cal G}_{ab} x^{a} x^{b} + \tilde{{\cal G}}_{ab} x^{a} \partial^b +\frac{i}{\omega} \tilde{\tilde{{\cal G}}} _{ab}\partial^a \partial^b,\cr
	{\cal Q}_U^\omega &=& \frac{i}{\omega}\left(1-\frac{\omega^2}{2}\left(x^-\right)^2\right) {\cal G}_{--} -i x^- \omega x^a {\cal G}_{-a}+x^- \tilde{\cal G}_{-a} {\partial}^{a}, \cr
	{\cal Q}_N^\omega&=&  x^- {\cal G}_{--} +    {\cal G}_{-a} x^a + \frac{2i}{\omega}\tilde{\cal G}_{-a} \partial^{a},
\end{eqnarray}
where ${\cal G}_{ab}$, $\tilde{{\cal G}}_{ab}$, and $\tilde{\tilde{{\cal G}}} _{ab}$ are integrals of the components of the Riemann tensor $R_{+\bar{a}+\bar{b}}$ evaluated on the geodesic -- see Supplementary Note  1:
\begin{equation}
    \label{Eq6}
	{\cal G}_{\bar{a}\bar{b}}= 
	\int _0^{\tau}\!\!\!d\tau R_{+\bar{a}+\bar{b}}\,,\quad
	\tilde{{\cal G}}_{\bar{a}\bar{b}}=\int _0^{\tau}\!\!\!d\tau {\cal G}_{\bar{a}\bar{b}}\,,\quad
	\tilde{\tilde{{\cal G}}}_{\bar{a}\bar{b}}= \int _0^{\tau}\!\!\!d\tau \tilde{{\cal G}}_{\bar{a}\bar{b}}\,,\quad
\end{equation}
where $\tau$ is the affine parameter on the geodesic.  {The distortions provided by \eqref{Eq5}is the solution to the exact quantum effective action and cannot be reproduced by motion of a point-like particle along a geodesic. They do not exist in flat spactime geometry, so they manifest a set of quantum effects in curved spacetime geometry.}  A similar approach can be used to find the wavefunction of a massive scalar particle. The physical degrees of  $U(1)$ gauge fields get corrected as if they were scalar fields, see Supplementary Note 2 and Supplementary Note 3. Supplementary Note 4 presents the operators in the Hilbert space that corresponds to these corrections. The following subsections show how these operators distort the physical information encoded in wave-packets traveling along couple of examples of null geodesics in the Solar system and around the Earth.

We now study the distortion operators in a couple of spacetime geometries, including de Sitter and Schwarzschild spacetime geometries. We first study the de Sitter and anti de Sitter space-time geometries because their symmetry allows one to immediately write down the components of the Riemann tensor in Fermi coordinates evaluated on the geodesic. \newline
\begin{figure}[t]
	\begin{center}
		\includegraphics[width=1\columnwidth]{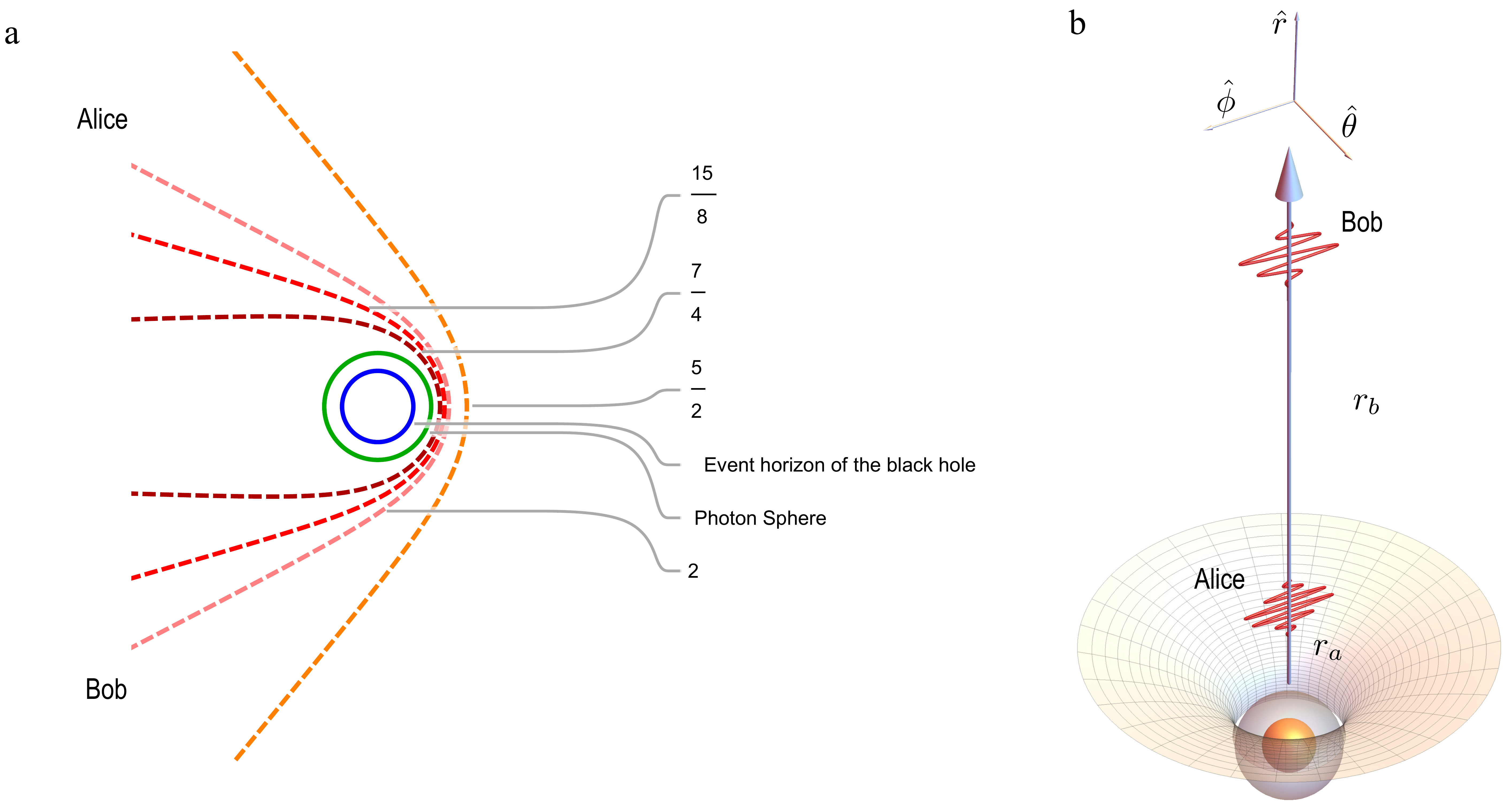}
		\caption[]{{\bf Propagation of wavepacket and associated null geodesics in the Schwarzschild spacetime geometry.} {\bf a} The null geodesics (dashed curves) for beams that propagate very close to the event horizon - we set the event horizon at 1. {\bf b} Schematic of wavepacket propagation radially in the Schwarzschild spacetime geometry. Alice and Bob are located at $r_a$ and $r_b$, respectively, while $\theta$ and $\phi$ are polar and azimuthal angles of the standard spherical coordinates.}
		\label{fig:fig2}
	\end{center}
\end{figure}
\textbf{de Sitter and anti de Sitter spacetime geometries:} The de Sitter and anti de Sitter space-times are maximally symmetric and the Riemann tensor at any given event in the space-time in any coordinates, including the Fermi coordinates, is given by  $R_{\mu\nu\mu'\nu'}= \Lambda (g_{\mu\mu'}g_{\nu\nu'}-g_{\mu\nu'}g_{\nu\mu'})$.  The value of $\Lambda$ determines different geometries: $\Lambda>0$ represents the de Sitter spacetime geometry; $\Lambda<0$ represents the anti de Sitter spacetime geometry; and $\Lambda=0$ is the Minkowki spacetime geometry. $R_{+-+-}=\Lambda$ is the only non-zero component for $R_{+\bar{a}+\bar{b}}$ evaluated on the geodesic, and thus the correction operators, Eq.~\eqref{Eq5} are,
\begin{equation}
	\label{Eq7}
	{\cal O}^\omega = 0,~
	{\cal Q}_U^\omega= \frac{i}{\omega}\left(1-\frac{\omega^2}{2}\left(x^-\right)^2\right) \Lambda x^+,~ 
	{\cal Q}_N^\omega=   \Lambda x^+ x^- \,.
\end{equation}
Let us consider a Gaussian wavepacket with normal distribution for $\omega$ around $\omega_0$ with the width of $\sigma$, i.e. $\psi_{\text{Alice}} =f^{(0)}(x^+,x^1,x^2) e^{i \omega_0 x^-} e^{-\frac{(\sigma x^-)^2}{2}}$. The validity of the perturbative solution demands that $|\Lambda| \ll \omega_0^2$, $|\Lambda|\ll \sigma^2$ and $\sigma \ll\omega_0$. The wavepacket after the propagation is ${\psi}_{\text{Bob}}\simeq (1+ \epsilon {\cal Q}^{\omega_0}_U+\epsilon {\cal Q}^{\omega_0}_{N})e^{i \omega_0 x^-}\,f^{(0)} e^{-\frac{(x^-)^2 \sigma^2}{2}}$.

${\cal Q}^{\omega_0}_{N}$ and ${\cal Q}^{\omega_0}_{U}$ change the wavepacket amplitude and phase, respectively. The maximum of $|{\cal Q}^{\omega_0}_{N} \psi_{\text{Alice}}|$ occurs at $x^-=\pm{1}/{\sigma}$. Requiring it to be  smaller than 1 yields  $T < \tau_A$ where $\tau_A= {(\sigma\sqrt{e})}/{|\Lambda|}$. $\tau_A$ is the maximum time that the wavepacket feels the curvature of the spacetime geometry and keeps its amplitude intact. ${\cal Q}^{\omega_0}_{U}$ alters the phase of the wavepacket. The maximum of $|{\cal Q}^{\omega_0}_{U} \psi_{\text{Alice}}|$ occurs at $x^-=\pm{\sqrt{2}}/{\sigma}$. Requiring it to be smaller than 1 results in $T <\tau_\phi$ where  $\tau_\phi={(\sigma\,\tau_A\,\sqrt{e})}/{(2\omega_0)}$. $\tau_\phi$ is the maximum time that a wavepacket can feel the curvature of the spacetime geometry and keep its phase intact. For $T>\tau_\phi$, information about phase is lost at perturbation.  {This may point to} a ``gravitational decoherence'', and its possible consequence on anti-de Sitter/conformal field theory correspondence
\cite{Maldacena:1997re} demands attention. $\tau_A$ represents the  amount of time of interaction with the curvature that the wavepacket can keep its amplitude intact. We observe that $\tau_\phi \ll \tau_A$.  So the phase changes sooner than the change in the amplitude.

\noindent\textbf{Schwarzschild spacetime geometry:} 
We choose the standard spherical coordinates $r,\theta,\phi$ where geodesics are extrema of,
\begin{equation}
    \label{Eq8}
	{\cal L} = -\left(1-\frac{m}{r}\right)\dot{t}^2+ \left(1-\frac{m}{r}\right)^{-1}\,\dot{r}^2+ r^2 (\dot{\theta}^2+ \sin^2 \theta \dot{\phi}^2)\,,
\end{equation}
and $m =2 G\,M_{\bullet}$ is the Schwarzschild radius, $M_{\bullet}$ is the mass of the blackhole, and $G$ is the gravitational constant -- the units are such that $m=1$. The components of the Riemann tensor in the Fermi-coordinates adapted to a general null geodesic of Schwarzschild spacetime geometry are derived in the Supplementary Note 5. Figure~\ref{fig:fig2}-{\bf a} shows several null geodesics that go very close to a blackhole. We first consider that the wavepacket  propagates along the radial direction, Figure~\ref{fig:fig2}-{\bf b},  where the only non-zero components of the Riemann tensor is $R_{+-+-} = -{1}/{r^3}$. This is the same component that appeared in the de Sitter spacetime geometry.  The radial geodesic has $l=0$, and its correction operators are ${\cal O}^\omega=0$, ${\cal Q}^\omega_U= -\frac{i}{2\omega}\left(1-\frac{\left(\omega x^-\right)^2}{2}\right)\left(\frac{1}{r_a^2}-\frac{1}{r^2}\right)$, and ${\cal Q}^\omega_N=-\frac{1}{2}\left(\frac{1}{r_a^2}-\frac{1}{r^2}\right) x^-$, where Alice is located at $r_a$. These correction terms do not contain derivatives of the spatial transverse coordinates. Thus, the Riemann tensor does not affect the spatial transverse profile of the wavepacket. This is due to the symmetry, as the radial geodesic inherits the static and spherical symmetry of the background. 

\begin{figure}[t]
	\begin{center}
		\includegraphics[width=1\columnwidth]{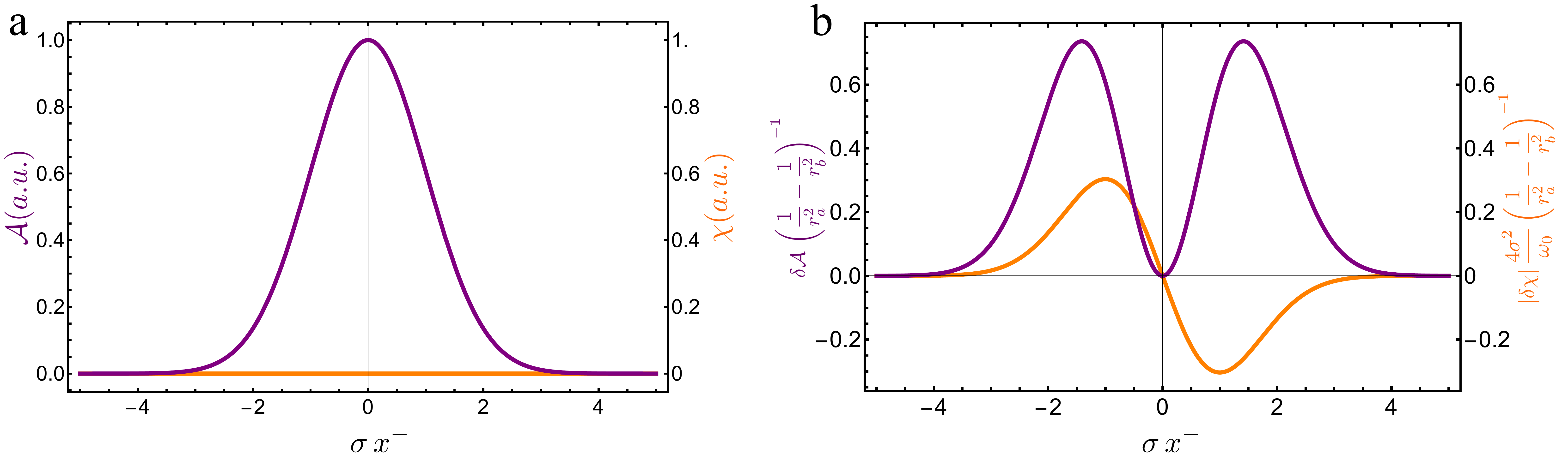}
		\caption[]{{\bf Propagation of Gaussian wavepacket along radial direction in the Schwarzschild spacetime geometry.} {\bf a} Amplitude (purple) and phase (orange) of the initial Gaussian beam at the sender (Alice) respectively presented by ${\cal A}(a.u.)$ and $\chi(a.u.)$. Their units, $a.u.$, are chosen such that both are normalized to their maximum values. {\bf b} The changes in the amplitude ($\delta{\cal A}$) and the phase ($\delta\chi$) of the Gaussian wavepacket around frequency $\omega_0$ with width of $\sigma$ transmitted over the radial null geodesic in the Schwarzschild geometry between $r_a$ and $r=r_b$, the geometry is shown in Fig.~\ref{fig:fig2}-{\bf b}. The units are chosen where the Schwarzschild radius and light speed are one.}
		\label{fig:fig3}
	\end{center}
\end{figure}

Figure \ref{fig:fig3}-\textbf{a} shows the amplitude and phase of a Gaussian time-bin signal. Figure \ref{fig:fig3}-\textbf{b} depicts the alterations in the amplitude and the phase; %
\begin{eqnarray}\label{Eq9}
	\delta {\cal A}&=&-\frac{\epsilon x^-}{2} \left(\frac{1}{r_a^2}-\frac{1}{r_b^2}\right)  e^{-\frac{\left(\sigma x^-\right)^2}{2}}+O\left(\epsilon^2\right) \,\\ \nonumber
	\delta {\chi}&=&\frac{i \omega_0}{4\sigma^2 }\left(\frac{1}{r_a^2}-\frac{1}{r_b^2}\right) \left(\sigma x^-\right)^2 e^{-\frac{\left(\sigma x^-\right)^2}{2}}+O\left(\epsilon^2\right) \,,
\end{eqnarray}
where $r_b$ is the location of Bob. The maximum alteration to the amplitude and phase of the Gaussian wavepacket occur at $x^-=\pm{1}/{\sigma}$ and $x^-=\pm{\sqrt{2}}/{\sigma}$, respectively. For a Gaussian wavepacket that propagates radially close to the Earth, the maximum alteration to the amplitude and phase are respectively $|\delta {\cal A}_{\text{max.}}|=\frac{1}{2 \sigma}\left(\frac{1}{r_a^2}-\frac{1}{r_b^2}\right)$ and $\delta {\chi}_{\text{max.}}=\frac{\omega_0\,m_\oplus}{2   \sigma^2}\left(\frac{1}{r_a^2}-\frac{1}{r_b^2}\right)$, where $m_\oplus$ is the Schwarzschild radius of Earth. 

For $\nu_0=456$~THz and  $\Delta\nu=1$~kHz, $\delta {\chi}_{\text{max.}}=0.10$~rad. Supplementary Note 8 provides further details on choosing these values.

As a final example, we examine the weak regime of gravity when the beam possesses well-defined transverse modes -- see Fig.~\ref{fig:fig2}-({\bf a}). The wavepacket carrying a well-defined transverse mode traverses through space and reaches to the minimum distance of $l$ to the central mass -- here, we assume $l$ is large.  We now consider a specific wavepacket, a Hermite-Gauss transverse mode $f_{\omega,p,\ell,q}^{(0)}(x^-,x^+,x^1,x^2)=e^{i \omega_0 x^-} e^{-\frac{(\sigma x^-)^2}{2}}\,\text{HG}_{m,n}(x^+,x^1,x^2)$ -- Hermite-Gauss modes are used to extend the communication alphabet beyond bits, i.e., 0 and 1~\cite{Mario:2014}. The longitudinal and frequency distributions are assumed to be Gaussian. When $\sigma$ is large, the dominant correction operator  is calculated to be:
\begin{eqnarray}
    \label{Eq10}
	\cal O &=&- \frac{9.5 i a^2}{z_R}\left(x^2-y^2\right) \,,
\end{eqnarray}
where $x={\sqrt{2} x_1}/{w(x^+)}$ and $y={\sqrt{2} x_2}/{w(x^+)}$ are dimensionless coordinates, $w(x^+)=w_0 \sqrt{1+ \left(\frac{x^+}{z_R}\right)^2}$ is the beam radius, $z_R=\frac{1}{2}{\omega}_0 w_0^2$ is the Rayleigh range, $a={r_a}/{l}$ is a scaling parameter, $w_0$ is the beam radius at Alice's position - see Supplementary Note 6 for more details. This operator contains coordinate parameters $x$ and $y$, and thus, alters both the amplitude and phase of the transverse modes upon propagation. The correction for the solar system, when Alice and Bob are at the mean Earth-Sun distance from the Sun and the wavepacket passes at $l=2R_\odot$, and for $z_R=2.8~\text{km}\times a^2 = 1.34\times 10^{9}~\text{m}$, remains perturbative and is given by $\epsilon\psi^{(1)}(x^\mu)= i\,0.10\,\left(x^2-y^2\right) \psi_{\text{Alice}}(x^\mu)|_{x^+=T}$.

\begin{figure*}[t]
	\begin{center}
		\includegraphics[width=2\columnwidth]{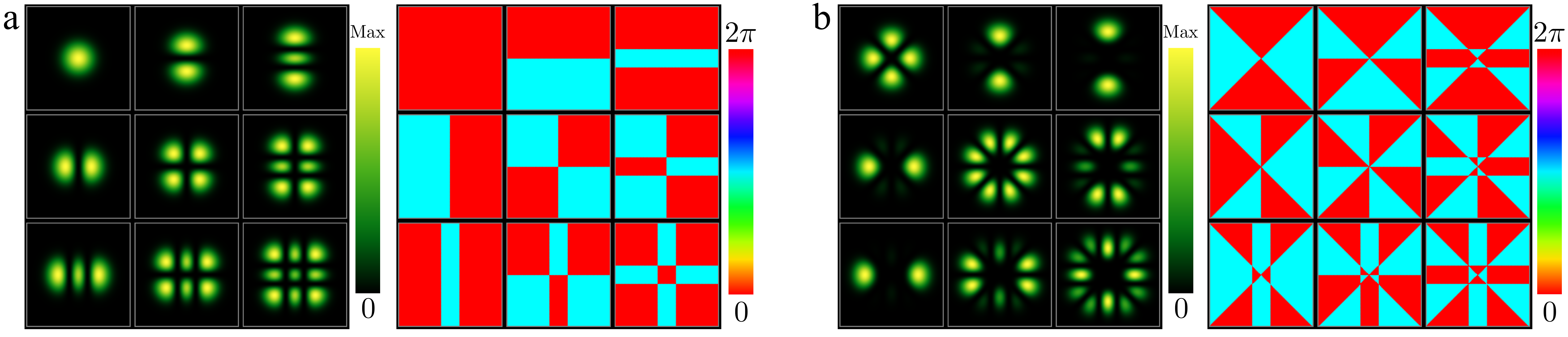}
		\caption[]{{\bf Propagation of Hermite-Gaussian modes in a weak gravitational field.}  Hermite-Gaussian mode $\text{HG}_{m,n}(x^+,x^1,x^2)$ is not shape invariant under propagation in the Schwarzschild spacetime geometry, and both the intensity and phase profiles modify. {\bf a} Intensity and phase distributions of the first 9 HG modes $m,n\in\{0,1,2\}$ prior to the free-space propagation. {\bf b} The intensity $|\epsilon\psi^{(1)}|^2$ and phase $\arg{\epsilon\psi^{(1)}}$ distributions of the corrected term, for the first 9 HG modes $m,n\in\{0,1,2\}$, after the propagation in a weak gravitational field.} The rows and columns are associated to $m\in\{0,1,2\}$ and $n\in\{0,1,2\}$, respectively. The dimensionless coordinates are used for these plots.
		\label{fig:fig4}
	\end{center}
\end{figure*}

The amplitude and phase of the mode $\psi^{(0)}(x^\mu)$ and the correction $\epsilon\psi^{(1)}(x^\mu)$ for a few Hermite-Gaussian modes are shown in Fig.~\ref{fig:fig4}. As seen, these alterations on the modes are considerable. For instance, it causes up to 12.2\% crosstalk between HG$_{0,3}$ and HG$_{0,1}$ modes. The crosstalk between mode $\cal{M}$ and mode $\cal{N}$ is given by  $\abs{\braket{\cal{N}}{\cal{M}}}^2$. Therefore, these alterations need to be accounted for when information is encrypted in the spatial modes. The action of curved spacetime geometry on the wavepacket is linear. Therefore, a target beam that does not possess the information can be used as a reference to monitor the distortion of the information carrier, and an active system can be employed for compensating the distortion in real-time -- in conjunction, results in retrieving the original information. Moreover, whenever $\epsilon\simeq1$, the higher-order terms of correction need to be considered. For instance, for $z_R< 28~\text{km}\times a^2$, the correction becomes larger than 1 and we need to take into account higher $\epsilon$ terms. Taking into account all the corrections is tantamount to knowing the Riemann tensor in whole of the spacetime geometry, a piece of knowledge which is not attainable. Thus, we tend to argue that once the perturbation breaks, in addition to known well-studied gravitational decoherence \cite{Pang:2016foq,Stefanov:2020kpw,Bassi:2017szd,Pang:2016foq,Penrose:1996cv,Joshi:2017fhz}, a decoherence occurs. We observe that, in addition to the known decoherence of a bipartite entangled system when each particle traverses through a different gravitational field gradient \cite{Joshi:2017fhz,Ralph1}, a coherent beam  decoheres when different segments of the spatial spread of the wave  experience  different tidal gravitational field gradients.  The phenomenon we are reporting also occurs for geodesics passing very close to the event horizon -- see Supplementary Note 7. 

Finally, it is noteworthy that photon pairs $\ket{\psi}_\text{entangled}$, e.g. entangled in spatial, frequency or temporal modes, would be affected by the curved spacetime geometry whenever they are shared between two parties, namely Alice and Bob. The final state of the entangled photon, indeed, is given by applying the non-local operators ${\cal U}=\left(1+\epsilon ({ \cal O}+ {\cal Q}_U + {\cal Q}_N) \right)$ onto the entangled states, $\left({\cal U}_A\otimes{\cal U}_B\right)\ket{\psi}_\text{entangled}$ - here, ${\cal U}_A$ and ${\cal U}_B$ are associated with the correction operators at Alice and Bob's places, respectively.

\section*{Conclusion} We have presented how the curvature of the spacetime geometry affects the propagation of an arbitrary wavepacket along a general geodesic in a general curved spacetime geometry. The effect is beyond classical general relativity, residing in the same category as Hawking radiation \cite{1974Natur.248...30H}. A set of linear operators are presented that encode the effect of the curvature. The corrections to the information carrier wavepacket are investigated in cases of de Sitter (anti de Sitter) and Schwarzschild spacetime geometries. It has been shown that the corrections accumulate over time and  distort the wavepacket. The gravitational distortion, therefore, needs to be accounted for in quantum communication performed over long distances in a curved spacetime geometry.

\noindent\textbf{Acknowledgments}
\noindent This work was supported by the High Throughput and Secure Networks Challenge Program at the National Research Council of Canada, the Canada Research Chairs (CRC) and Canada First Research Excellence Fund (CFREF) Program, and Joint Centre for Extreme Photonics (JCEP). The authors would like to thank Alicia Sit, Benjamin Sussman, Khabat Heshami, Thomas Jennewein, Christoph Simon, Mathias Blau, Ida Zadeh and Loriano Bonora for fruitful discussions and thoughtful feedback, \textcolor{black}{and Haorong Wu for the email correspondence.} Q.E. would like to thank Fernando Quevedo and Atish Dabholkar for the nice hospitality in ICTP where part of the work was conducted. 
\vspace{1 EM}

\noindent\textbf{Author contributions}
\noindent
Q.E. and E.K. conceived the idea and developed the theoretical framework. E.C. checked and confirmed the computation. All authors contributed to the manuscript preparation.
\vspace{1 EM}

\noindent\textbf{Data availability} The authors declare that the data supporting the findings of this study are available within the paper and its supplementary information file. 
\vspace{1 EM}

\noindent\textbf{Competing interests} The authors declare no competing interests.
\vspace{1 EM}

\noindent\textbf{Author Information}
\noindent Correspondence and requests for materials should be addressed to Q.E. (qexirifa@uottawa.ca) or E.K. (ekarimi@uottawa.ca).
\vspace{1 EM}

\renewcommand{\figurename}{\textbf{Supplementary Figure}}
\setcounter{figure}{0} \renewcommand{\thefigure}{\textbf{S{\arabic{figure}}}}
\setcounter{table}{0} \renewcommand{\thetable}{S\arabic{table}}
\setcounter{section}{0} \renewcommand{\thesection}{S\arabic{section}}
\setcounter{equation}{0} \renewcommand{\theequation}{S\arabic{equation}}
\onecolumngrid
\setcounter{page}{1}

\vspace{1 EM}
\newpage

\clearpage

\title{Supplementary Information for Towards Communication in a Curved Spacetime Geometry}
\author{Qasem Exirifard}
\author{Eric Culf}
\author{Ebrahim Karimi}
\affiliation{Department of Physics, University of Ottawa, 25 Templeton St., Ottawa, Ontario, K1N 6N5 Canada; ekarimi@uottawa.ca, qexirifa@uottawa.ca}
\maketitle
\onecolumngrid
\section*{\textbf{\large{Supplementary Note 1: Leading order correction term}}}\label{SI1}
We consider the Fourier integral representation of $\psi^{(1)}$ in terms of $x^-$: 
\begin{equation}\label{PSI1F1Def}
	\psi^{(1)} =\int d\omega  f^{(1)}_\omega(x^+,x^a) e^{i\omega x^-}\,,
\end{equation}
where  $f^{(1)}_\omega$ is the correction to the structure function for the frequency $\omega$. Substituting \eqref{PSI1F1Def} and $\psi^{(0)} =\int d\omega  f^{(0)}_\omega(x^+,x^a) e^{i\omega x^-}$ in $\Box^{(0)} \psi^{(1)}  =  \, +\omega^2 R_{+\bar{a}+\bar{b}}~x^{\bar{a}} x^{\bar{b}}\psi^{(0)} $ yields,
\begin{eqnarray}\label{Papneeded1}
	\int\!d\omega e^{i\omega x^-}  \left(2i \omega \partial_+ + \nabla^2_{\perp}\right) f^{(1)}_\omega = \int\!d\omega \omega^2  R_{+\bar{a}+\bar{b}}~x^{\bar{a}} x^{\bar{b}}f^{(0)}_\omega e^{i\omega x^-}.
\end{eqnarray}
We can use $x^- e^{i\omega x^-}=-i\partial_\omega e^{i\omega x^-}$, and integration by parts to simplify the r.h.s of the above equation to,
\begin{eqnarray}\label{Papneeded2}
	\int\!d\omega \left[-\left(\omega^2 f^{(0)}_\omega\right)'' R_{+-+-}+ 2i\left(\omega^2 f^{(0)}_\omega\right)'x^a R_{+-+a}+ \omega^2 f^{(0)}_\omega  x^a x^b R_{+a+b}\right]  e^{i\omega x^-}\,,
\end{eqnarray}
and then obtain,
\begin{equation}\label{EqPara-1}
	\left(2 i \omega \partial_++ \nabla^2_{\!\!\perp}\right) f^{(1)}_\omega  =  \, -(\omega^2 f^{(0)}_\omega)'' R_{+-+-}+ 2i (\omega^2 f^{(0)}_\omega)'x^a R_{+-+a}+\omega^2 f^{(0)}_\omega  x^a x^b R_{+a+b} \,,
\end{equation}
where $()'$ stands for the derivatives w.r.t. $\omega$. We would like to solve 	\eqref{EqPara-1} for all $f^{(0)}$ that solves
\begin{equation}
	\label{f0equationAux}
	(\nabla_{\!\!\perp}^2 + 2 i \omega\partial_+) f^{(0)}=0.
\end{equation}
So $f^{(0)}_\omega$ can be expanded in  the Hermite-Gaussian basis or the Laguerre-Gaussian basis. In the  Hermite-Gaussian basis, one can write
\begin{equation}
	\label{S6}
	f^{(0)}_\omega(x^+,x^a)= \sum_{l,n} c_{ln}(\omega) f_{ln,\omega}(x^+,x^a), 
\end{equation}
where $c_{ln}(\omega) $ is the coefficient of the expansion in the Hermite-Gaussian basis. This implies 
\begin{equation}\label{dOmegaf0}
	\partial_\omega  f^{(0)} = \sum_{l,n} \left( f_{ln,\omega}\partial_\omega c_{ln}+ c_{ln} \partial_\omega  f_{ln,\omega} \right).
\end{equation}
We consider field configurations wherein 
\begin{equation}\label{Condition}
	|f_{ln,\omega}\partial_\omega c_{ln}| \gg |c_{ln} \partial_\omega  f_{ln,\omega}|.
\end{equation}
which means that the frequency dependency of the structure function is mainly given by the coefficients of $c_{ln}$. This means that we are not considering white noise, and we are considering a wavepacket that has a sufficiently sharp Gaussian peak in the frequency spectrum. We are considering configuration that have sharp peak around the mean frequency.  Because of  \eqref{Condition}, we can approximate \eqref{dOmegaf0} to
\begin{equation}
	\partial_\omega  f^{0} = \sum_{l,n} f_{ln,\omega}\partial_\omega c_{ln},
\end{equation}
which can be utilized to write
\begin{equation*}
	(\nabla_\perp^2 + 2 i\omega \partial_+) \partial_\omega  f^{0} =\sum_{l,n} \partial_\omega c_{ln} (\nabla_\perp^2 + 2 i\omega \partial_+) f_{ln,\omega} .
\end{equation*}
Due to \eqref{f0equationAux} and \eqref{S6}, it holds,
\begin{equation*}
	(\nabla_{\!\!\perp}^2 + 2 i \omega\partial_+) f_{ln,\omega} = 0
\end{equation*} 
Therefore,
\begin{eqnarray}
	\label{ConditionAux0}
	(\nabla_\perp^2 + 2 i\omega \partial_+) \partial_\omega  f^{0} &=&0 .
\end{eqnarray}
Utilizing the same method, eq. \eqref{Condition}, implies that all derivatives of $f^{(0)}$ with respect to $\omega$ solve \eqref{f0equationAux}. The right-hand-side of  \eqref{EqPara-1}, therefore, includes terms that vanish by the same operator which appears in the left-hand-side of \eqref{EqPara-1}. This resembles resonance where the source oscillates with the frequency of the system. In the following we, show that, similar to the resonance, the corrections accumulate over time.
To solve \eqref{EqPara-1}, along the geodesic $\gamma$ with $\dot{\gamma}^\mu=\hat{x}^+$, we define:
\begin{subequations}\label{eq21aux}
	\begin{eqnarray}
		{\cal G}_{\bar{a}\bar{b}}&=& 
		\int _0^{x^+}\!\!\!d\xi^+ ~ R_{+\bar{a}+\bar{b}}(\xi^+)\,,\\
		\tilde{{\cal G}}_{\bar{a}\bar{b}}&=& \int _0^{x^+}\!\!\!d\xi^+ ~{\cal G}_{\bar{a}\bar{b}}(\xi^+)\,,\\
		\tilde{\tilde{{\cal G}}}_{\bar{a}\bar{b}}&=& \int _0^{x^+}\!\!\!d\xi^+ ~\tilde{{\cal G}}_{\bar{a}\bar{b}}(\xi^+)\,,
	\end{eqnarray}
\end{subequations}
where $\xi^+$ is the affine parameter on the geodesic.  
Next write $f^{(1)}$ by:
\begin{equation}\label{fansatz}
	f^{(1)}_\omega= \tilde{f}^{(1)}_\omega- \frac{\left(\omega^2 f^{(0)}_\omega\right)''}{2 i\omega} {\cal G}_{--} + \frac{\left(\omega^2 f^{(0)}_\omega\right)'}{\omega} x^a {\cal G}_{-a}- \frac{i\omega}{2}  {\cal G}_{ab} x^{a} x^{b} f^{(0)}_\omega\,,
\end{equation}
where $\tilde{f}^{(1)}$ is an arbitrary function of $x^+$ and $x^a$. Utilizing \eqref{fansatz} in \eqref{EqPara-1} yields:  
\begin{eqnarray}\label{tildef}
	\left(2 i \omega \partial_++ \nabla^2_{\!\!\perp}\right) \tilde{f}^{(1)}_\omega= 2 i  \omega {\cal G}_{ab}x^a  \partial^b f^{(0)}_\omega+ \frac{i\omega}{2} {\cal G}_{ab} f^{(0)}_\omega\nabla^2_\perp\left(x^a x^b\right)-\frac{2}{\omega}  {\cal G}_{-a}\partial^a \left(\omega^2 f^{(0)}_\omega\right)' 
\end{eqnarray}
We notice that the $++$ component of the Ricci tensor along the geodesic holds:
\begin{equation}\label{RicciPlusPlus}
	\eta^{\mu\nu} R_{+\mu+\nu} = R_{+a+b}\eta^{ab}=0\,,
\end{equation}
which implies $\delta^{ab} {\cal{G}}_{ab}=0$ and simplifies \eqref{tildef} to,
\begin{eqnarray}\label{ftildeEq}
	\left(2 i \omega \partial_++ \nabla^2_{\!\!\perp}\right) \tilde{f}^{(1)}_\omega  =  \, 2i  \omega {\cal G}_{ab} x^{a}  \partial^b f^{(0)}_\omega-\frac{2}{\omega}  {\cal G}_{-a}\partial^a \left(\omega^2 f^{(0)}_\omega\right)'\,.
\end{eqnarray}
Now let
\begin{equation}\label{ft}
	\tilde{f}^{(1)}_\omega=\tilde{\tilde{f}}^{(1)}_\omega+  \tilde{{\cal G}}_{ab} x^{a} \partial^b f^{(0)}_\omega+ \frac{i}{\omega^2}  	\tilde{\cal G}_{-a}\partial^a \left(\omega^2  f^{(0)}_\omega\right)'\,,
\end{equation}
where $\tilde{\tilde{f}}^{(1)}_\omega$ is an arbitrary function of $x^+$ and $x^a$. Utilizing \eqref{ft} in \eqref{ftildeEq} yields:
\begin{equation}
	\left(2 i \omega \partial_++ \nabla^2_{\!\!\perp}\right)\tilde{\tilde{f}}^{(1)}_\omega = -2 \tilde{\cal G}_{ab} \partial^{a}\partial^b f^{(0)}_\omega\,,
\end{equation} 
where we have noticed that derivatives of solutions to homogeneous differential equations are also solutions. The solution for $\tilde{\tilde{f}}^{(1)}$ is given by,
\begin{equation}\label{ftt}
	\tilde{\tilde{f}}^{(1)}_\omega = \frac{i}{\omega} \tilde{\tilde{{\cal G}}} _{ab}\partial^a \partial^b f^{(0)}_\omega \,.
\end{equation}
Combining \eqref{fansatz},\eqref{ft} and \eqref{ftt} gives:
\begin{eqnarray}\label{f1exact}
	f^{(1)}_\omega=\left(- \frac{i\omega}{2}  {\cal G}_{ab} x^{a} x^{b} + \tilde{{\cal G}}_{ab} x^{a} \partial^b +\frac{i}{\omega} \tilde{\tilde{{\cal G}}} _{ab}\partial^a \partial^b\right) f^{(0)}_\omega+\left(\frac{ i}{2\omega} {\cal G}_{--} \left(\omega^2 f^{(0)}_\omega\right)'' +\frac{1}{\omega} x^a {\cal G}_{-a}\left(\omega^2 f^{(0)}_\omega\right)' + \frac{i}{\omega^2} \tilde{\cal G}_{-a} \partial^a \left(\omega^2 f^{(0)}_\omega\right)'\right)
\end{eqnarray}
which we also directly have checked that solves \eqref{EqPara-1}. Let us define:
\begin{eqnarray}\label{OO}
	{\cal O}^\omega &=& - \frac{i\omega}{2}  {\cal G}_{ab} x^{a} x^{b} + \tilde{{\cal G}}_{ab} x^{a} \partial^b +\frac{i}{\omega} \tilde{\tilde{{\cal G}}} _{ab}\partial^a \partial^b\,,\\ \label{QO}    
	{\cal Q}[f^{(0)}_\omega]&=& \frac{ i}{2\omega} {\cal G}_{--} \left(\omega^2 f^{(0)}_\omega\right)'' +\frac{1}{\omega} x^a {\cal G}_{-a}\left(\omega^2 f^{(0)}_\omega\right)' + \frac{i}{\omega^2} \tilde{\cal G}_{-a} \partial^a \left(\omega^2 f^{(0)}_\omega\right)'\,.
\end{eqnarray}
Therefore,
\begin{eqnarray}\label{PsiCompact}
	\psi&=&\psi^{(0)}+\epsilon \psi^{(1)} + O(\epsilon^2)\\ \nonumber
	&=&\int d\omega e^{i\omega x^-} \left(f^{(0)}_\omega+\epsilon { \cal O}^\omega f^{(0)}_\omega+\epsilon {\cal Q}[f^{(0)}_\omega]\right)+O(\epsilon^2) \,,
\end{eqnarray}
Integration by parts can be employed to rewrite \eqref{PsiCompact} to: 
\begin{eqnarray}\label{PsiCompact2}
	\psi(x^\mu)&=&\int d\omega~ e^{i\omega x^-} \left(1+\epsilon { \cal O}^\omega+\epsilon {\cal Q}_U^\omega +\epsilon {\cal Q}_N^\omega \right)f^{(0)}_\omega+O(\epsilon^2)  \,,
\end{eqnarray}
where 
\begin{eqnarray}\label{OQU}
	{\cal Q}_U^\omega &=& \frac{i}{\omega}\left(1-\frac{\omega^2}{2}\left(x^-\right)^2\right) {\cal G}_{--} -i x^- \omega x^a {\cal G}_{-a}+x^- \tilde{\cal G}_{-a} {\partial}^{a}\,, \\ \label{OQN}
	{\cal Q}_N^\omega&=&  x^- {\cal G}_{--} +    {\cal G}_{-a} x^a + \frac{2i}{\omega}\tilde{\cal G}_{-a} \partial^{a}\,.
\end{eqnarray}

\section*{\textbf{\large{Supplementary Note 2: Corrections to U(1) gauge field}}}
\label{U1section}
We would like to compute the correction due to the curvature of the spacetime geometry to an arbitrary structured photon propagating along a general null geodesic in a given curved spacetime geometry.   We start from the action of a $U(1)$ gauge field in a general curved spacetime endowed by  metric $g_{\mu\nu}$ which is given by,
\begin{eqnarray}
	\label{SA}
	S[{A}_\mu] &=& -\frac{1}{4} \int\!\!d^4\!x\,g^{\mu\mu'\nu\nu'} F_{\mu\nu} F_{\mu'\nu'}\,,\\
	g^{\mu\mu'\nu\nu'}& =&\frac{1}{2}\sqrt{-\det g} (g^{\mu\mu'}g^{\nu\nu'}-g^{\mu\nu'}g^{\nu\mu'}) \,,
\end{eqnarray}
where $F_{\mu\nu}$ is the field strength of the $U(1)$ connection:
$F_{\mu\nu} = \partial_\mu {A}_\nu-\partial_\nu {A}_\mu$.
The functional variation of the action with respect to the gauge field gives the its equation of motion:
\begin{equation}\label{U1Eq}
	\partial_\mu \left(g^{\mu\mu'\nu\nu'} F_{\mu'\nu'}\right)=0\,.
\end{equation}
We would like to study perturbative solutions for the $\epsilon$ expansion of the metric. To this aim, we write an expansion series for the gauge field and $g^{\mu\mu'\nu\nu'}$: 
\begin{eqnarray}\label{AE}
	A_{\mu} &=& A^{(0)}_\mu + \epsilon A^{(1)}_\mu   + O(\epsilon^2) \,,\\
	\label{AE2}
	g^{\mu\mu'\nu\nu'} &=& g^{(0) \mu\mu'\nu\nu'}+ \epsilon g^{(1)\mu\mu'\nu\nu'} + O(\epsilon^2) \,.
\end{eqnarray} 
Utilizing \eqref{AE} and \eqref{AE2} in  \eqref{U1Eq}  and keeping the zero order yields
\begin{eqnarray}
	\label{S31}
	\partial_\mu\left(g^{(0)\mu\mu'\nu\nu'} F_{\mu'\nu'}^{(0)}\right) &=& 0 \to \nonumber\\
	\left(\eta_{\mu\nu}\Box^{(0)}  - \partial_\mu \partial_\nu\right) A^{(0)\nu} &=&0\,,
\end{eqnarray}
where from here forward $\eta^{\mu\nu}$ is utilized to move up or down the indices:  $A^{(0)\nu}=\eta^{\nu\lambda}A^{(0)}_{\lambda}$. We choose the Lorentz gauge, 
$\partial_\nu A^{(0)\nu} =0$,
that simplifies the equation for $A^{(0)}$ to,
\begin{eqnarray}\label{BoxA0}
	\Box^{(0)} A^{(0)}_\mu = \left(2 \partial_+ \partial_- + \nabla^2_{\perp}\right) A^{(0)}_\mu=0\,.
\end{eqnarray}
We are interested in the structured photon whose field configuration is given by,
\begin{equation}\label{ParaxialApproxPhoton}
	A^{(0)}_\mu = f^{(0)}_\mu(x^+,x^a) e^{i\omega x^-} \,,
\end{equation}
and the structure function, $f^{(0)}$, is very slowly varying: 
$|\partial_\mu f^{(0)}_\nu|\ll \omega |f^{(0)}_\nu|$. 
The Lorentz gauge then implies:
\begin{eqnarray}
	\label{f0zero}
	f^{(0)}_+&=&0\,,\\
	\partial_+ f_-^{(0)} +\partial^a f^{(0)}_a&=&0\,,
\end{eqnarray}
that  we choose to solve for $ f_-^{(0)}$. This leaves $f_a^{(0)}$ as the physical modes and maps \eqref{BoxA0} to,
\begin{eqnarray}\label{BoxAQM}
	\Box^{(0)} A^{(0)}_a = (2 i \omega \partial_+  + \nabla^2_{\perp}) A^{(0)}_a=0\,.
\end{eqnarray}
In the plane wave approximation where $ \nabla^2_{\perp} A^{(0)}_a$ vanishes, it holds,  $A^{(0)}_a = C_a e^{i\omega x^-}$. Here, $C_a$ are constants encoding the linear polarisation of the photon -- this is in accordance with the presentation used in~\cite{Goto}, where $A_1$ and $A_2$ are the transverse modes. They can be understood as the distribution of the polarisation of the photon.
Utilizing \eqref{AE} and \eqref{AE2} in  \eqref{U1Eq}  and keeping the linear terms in $\epsilon$ yield
\begin{eqnarray}
	\partial_\mu\left(g^{(0)\mu\mu'\nu\nu'} F_{\mu'\nu'}^{(1)}\right) &=&- \partial_\mu\left(g^{(1)\mu\mu'\nu\nu'} F_{\mu'\nu'}^{(0)}\right) \to \nonumber\\
	\label{EQAAux0}
	\Box^{(0)} A^{(1)\mu} &=& -\partial_\nu\left( g^{(1)\mu\mu'\nu\nu'} F^{(0)}_{\mu'\nu'}\right)\,.
\end{eqnarray}
where it is noted that the zero term in $\epsilon$ vanishes due \eqref{S31}, and  the Lorentz gauge is used too. This  the equation of motion for 	$A_\mu^{(1)}$.
We are interested in the solutions where $|\partial_\lambda g^{(1)\mu\nu}|\ll \omega |g^{(1)\mu\nu}|$. Therefore, $\partial_-$ in the rhs of \eqref{EQAAux0} does not act on the metric's components:
\begin{equation}\label{EQAAux10}
	\Box^{(0)} A^{(1)\mu}   = -g^{(1)\mu\mu'\nu\nu'}  \partial_\nu F^{(0)}_{\mu'\nu'}\,,
\end{equation}
As the beam's shape is changing very slowly compared to $\omega$, we should keep only $\partial_-$ in the rhs of \eqref{EQAAux10}:
\begin{equation}\label{EQAAux1}
	\Box^{(0)} A^{(1)\mu}   = -\omega^2\left(g^{(1)--\mu\alpha}-g^{(1)-\alpha\mu-}\right)  A_\alpha^{(0)}\,,
\end{equation}
wherein $\partial_-^2 A_\alpha^{(0)} = -\frac{\omega^2}{c^2} A_\alpha^{(0)} $ is utilized. Expressing \eqref{EQAAux1} in term of the components of the metric 
and utilizing  \eqref{f0zero} 
returns:  
\begin{eqnarray}
	\Box^{(0)} A^{(1)}_+ &=& 0\,,\\  
	\Box^{(0)} A^{(1)}_a &=&\delta\!g^{--} A^{(0)}_a\,,
\end{eqnarray}
the first of which is solved by  $A_+^{(1)}=0$. Recalling the Lorentz gauge, we note that $A^{(1)}_-$ can be expressed in term of $A^{(1)}_a$ so we have spared writing its equation. Utilizing the metric expansion series then gives,
\begin{eqnarray}\label{AEq}
	\Box^{(0)} A^{(1)}_i  =  \, \omega^2 R_{+\bar{a}+\bar{b}}~x^{\bar{a}} x^{\bar{b}}A^{(0)}_i  \,,
\end{eqnarray}
which  is identical to the equation for a scalar field.  So its solution can be straight fully derived from  \eqref{PsiCompact}. For the zero-order solution of,
\begin{equation}
	A^{(0)}_i = \int d\omega  A^{(0)}_i(\omega, x^+,x^a) e^{i \omega x^-}+ \text{c.c.} 
\end{equation}
The corrected solution is given by,
\begin{eqnarray}
	A_i&=& \int\!\!\! d\omega e^{i\omega x^-} \left(1+ \epsilon{\cal O}+  \epsilon{\cal Q}_U+ \epsilon {\cal Q}_N\right) A_i^{(0)} (\omega, x^+,x^a)+ O(\epsilon^2)\,,
\end{eqnarray}
where ${\cal O}$ and ${\cal Q}$ are defined in \eqref{OO} and \eqref{QO}. This concludes the correction for a $U(1)$ field.

\section*{\textbf{\large{Supplementary Note 3:  Corrections to massive scalar field}}}
\label{Massivescalar}
Alice and Bob can communicate by exchanging wavepackets made of massive particles.  We use the  Fermi Coordinates adapted to the time-like geodesic, given by Manasse and Misner~\cite{Manasse:1963zz}:
\begin{eqnarray}\label{RFM}
	ds^2 = + c^2 dt^2 \left(- 1+ R_{0l0m} x^l x^m\right) + \frac{2}{3} R_{0 l i m} x^l x^m dt dx^i+ dx^i dx^j \left(\delta_{ij}+ \frac{1}{3} R_{iljm} x^l x^m\right) +O(x^l x^m x^n)\,,
\end{eqnarray}
where $R_{\mu\alpha\beta\nu}$ represents the components of the Riemann tensor computed along the time-like geodesics of the packet. The components of the Riemann tensor in \eqref{RFM} are functions of time, and time is the proper time as measured in the rest frame of the massive wavepacket. We use the Greek alphabet $\mu$  from $\{0,1,2,3\}$ where $0$ stands for time. We use Latin indices to show the directions. The generally covariant form of a wave equation of a massive scalar particle $\phi$ with rest mass of $m_0$ is,
\begin{equation}\label{PsiM}
	\left(\Box- m_0^2\right) \phi =  \partial_\mu\left( g^{\mu\nu} \partial_\nu \phi\right) +\partial_\nu\phi g^{\mu\nu}\partial_\mu \ln(\sqrt{-\det g}) - m_0^2 \phi=0 \,.
\end{equation}
Note that we set the units such that $c=1$ and $\hbar=1$. Utilizing  the $\epsilon$ expansion series for the metric then
%
yields,
\begin{eqnarray}\label{massivePsiEq}
	\left(\Box^{(0)}- m_0^2\right) \phi &=&  -\epsilon \partial_\mu(\delta\!g^{\mu\nu} \partial_\nu \phi)-\epsilon \eta^{\mu\nu}\partial_\nu\phi \partial_\mu \delta\!g  + O(\epsilon^2)\,,
\end{eqnarray}
where 
$\Box^{(0)} = \eta^{\mu\nu}\partial_\mu \partial_\nu= -\partial_t^2 + |\nabla|^2$. 
We would like to solve \eqref{massivePsiEq} in a perturbative fashion. In so doing we consider:
\begin{equation}\label{psimassiveexpansion}
	\phi =f^{(0)} e^{i\omega t}+ \epsilon f^{(1)} e^{i \omega t} + O(\epsilon^2)\,,
\end{equation}
where $\omega:= m_0$ is understood. Next assume that 
$|\partial_\mu f|  \ll \omega |f| $
which  gives:
\begin{eqnarray}\label{Box0Massive}
	\left(-\frac{\hbar^2}{2m_0} \nabla^2 - i\hbar \partial_t \right) f^{(0)} &=& 0  \,,\\
	\label{Psi1MassiveEq}
	\left(-\frac{\hbar^2}{2m_0} \nabla^2 - i\hbar \partial_t \right) f^{(1)} &= & {\omega^2}\, R_{0l0m}(t) x^l x^m f^{(0)}  \,.
\end{eqnarray}
Eq. \eqref{Box0Massive} is the Schr\"odinger equation in $3+1$ dimensions.  Notice that \eqref{Psi1MassiveEq} can be directly obtained from considering the Schr\"odinger equation in the approximation of Newtonian potential associated to $g^{(1)}_{00}$.  The equation for $f^{(1)}$ can be mapped to a generalization of  \eqref{EqPara-1} to five dimensions  where the components of the  Riemann tensor in $5D$ vanishes for the $x^-$ transfer direction, and $t=x^+$. This mapping gives:
\begin{equation}\label{F1Godm}
	f^{(1)} =  \left(- \frac{i\omega}{2}  {\cal G}_{ab} x^{a} x^{b} + \tilde{{\cal G}}_{ab} x^{a} \partial^b +\frac{i}{\omega} \tilde{\tilde{{\cal G}}} _{ab}\partial^a \partial^b\right) f^{(0)},
\end{equation}
where 
\begin{eqnarray}
	\label{Gmassive}
	{\cal G}_{ab}&=& 
	\int _0^{T}\!\!\!d\tau ~ R_{0a0b}(\tau)\,,\\
	\tilde{{\cal G}}_{ab}&=& \int _0^{T}\!\!\!d\tau ~{\cal G}_{ab}(\tau)\,,\\
	\tilde{\tilde{{\cal G}}}_{ab}&=& \int _0^{T}\!\!\!d\tau ~\tilde{{\cal G}}_{ab}(\tau)\,,
\end{eqnarray}
where $\tau$ is an affine parameter and $a,b \in\{1,2,3\}$, and Bob gets the message $T$ seconds ($T$ in the rest frame of the particle) after Alice writes it. We directly have checked that \eqref{F1Godm} solves \eqref{Psi1MassiveEq} where we notice $R_{0a0}^{~~~a}=0$. 
We again notice that if we represent,
\begin{eqnarray}
	\phi_{\text{Alice}} &=& f^{(0)} e^{i\omega t}\,,\\
	\phi_{\text{Bob}} &=& \left(f^{(0)} +\epsilon f^{(1)}\right) e^{i\omega t}\,,
\end{eqnarray}
then,
\begin{equation}
	\phi_{\text{Bob}} = {\cal O}\, \phi_{\text{Alice}}\,,
\end{equation}
where the spacetime operator $\cal O$ depends on the spacetime geometry and the geodesic: 
\begin{equation}
	\label{OOmassive}
	{\cal O} = 1- \frac{i\omega}{2}  {\cal G}_{ab} x^{a} x^{b} + \tilde{{\cal G}}_{ab} x^{a} \partial^b +\frac{i}{\omega} \tilde{\tilde{{\cal G}}} _{ab}\partial^a \partial^b \,.
\end{equation}
This concludes the correction for a massive scalar.
\section*{\textbf{\large{Supplementary Note 4:  Corrections in the Hilbert space}}}
\label{Quantum}
We consider the Hilbert space for the scalar field $\psi$. We write an $\epsilon$ expansion series for the state $\ket{\psi}$:
\begin{equation}
	\label{correctedpsiHilbert}
	\ket{\psi}= \ket{\psi^{(0)}} + \epsilon \ket{\psi^{(1)}} + O(\epsilon^2)
\end{equation}
Let us consider:
\begin{equation}
	\label{XPsi}
	\braket{x}{\psi}= \braket{x}{\psi^{(0)}} + \epsilon \braket{x}{\psi^{(1)}} + O(\epsilon^2) = \psi^{(0)} +\epsilon \psi^{(1)}+ O(\epsilon^2)\,,
\end{equation}
where $\psi^{(0)}$ is,
\begin{equation}
	\psi^{(0)} = \int d\omega  f^{(0)}_\omega(x^+,x^a) e^{i \omega x^-}\,,
\end{equation}
and $\psi^{(1)}$ is presented in \eqref{PSI1F1Def}:
\begin{equation}
	\psi^{(1)} =\int d\omega  f^{(1)}_\omega(x^+,x^a) e^{i\omega x^-}\,.
\end{equation}
We notice that the equation that governs the dynamics of $f^{(0)}_\omega$ \eqref{f0equationAux}, which is also the paraxial approximation of the Helmholtz equation, is isomorphic to the Schr\"odinger equation. In fact 
$m_{eff} ={\hbar \omega}$ maps \eqref{f0equationAux} to the Schr\"odinger   equation for a free particle of mass $m_{eff}$ in $2+1$ dimensions of $(x^+, x^a)$: 
\begin{equation}
	-\frac{\hbar^2}{2m_\text{eff}} \nabla_{\!\!\perp}^2 f^{(0)}_\omega = i\hbar \partial_{x^+} f^{(0)}_\omega\,.
\end{equation}
This isomorphism implies that $x^+$ in the Hilbert space should be treated as a time-variable rather than an operator. The Hamiltonian associated to $x^+$ is:
\begin{equation}
	H = \frac{P^2}{2 m_\text{eff}}= \frac{1}{2\omega}(P^2_1+P^2_2)\,,
\end{equation}
where $P_1$ and $P_2$ are the momentum operators in the transverse directions $x^1$ and $x^2$. We choose a basis of the Hamiltonian to represent $f^{(0)}_\omega(x)$:
\begin{equation}
	\psi^{(0)} = \int d\omega e^{i\omega x^-} \sum_{s} C_{s,\omega}^{(0)} \exp\left(-\frac{i E_s x^+}{\hbar}\right) f_{s}(x^1,x^2)\,,
\end{equation}
where $f_{s}$ is an eigenstate of the Hamiltonian with energy of $E_s$. This allows us to write,
\begin{equation}
	\ket{\psi^{(0)}} = \int d\omega e^{i\omega x^-} \sum_{s} C_{s,\omega}^{(0)} \exp\left(-\frac{i E_s x^+}{\hbar}\right) \bm{a}^\dagger_{\omega,s}\ket{0}\,,
\end{equation}
where $\bm{a}^\dagger_{\omega,s}$ is the creation operator for a mode with frequency $\omega$ and structure $s$. $\ket{0}$ is the vacuum state in the Fermi coordinates for $\epsilon=0$ (adapted to the null geodesic). Eq. \eqref{PsiCompact2} presents the value for \eqref{XPsi}:
\begin{equation}
	\braket{x}{\psi}=\int d\omega~ e^{i\omega x^-} (1+\epsilon { \cal O}^\omega+\epsilon {\cal Q}_U^\omega +\epsilon {\cal Q}_N^\omega )\sum_{s} C_{s,\omega}^{(0)} \exp\left(-\frac{i E_s x^+}{\hbar}\right) f_{s}+O(\epsilon^2)  \,,
\end{equation}
In order to write $\ket{\psi}$, we first write the quantum analog of \eqref{OO},\eqref{OQU} and \eqref{OQN} by changing $x^a\to \bm{X}^a$ and $\partial_a \to \frac{i \bm{P}_a}{\hbar}$: 
\begin{eqnarray}\label{QUOO}
	\bm{{\cal O}}^\omega &=&- \frac{i\omega}{2}  {\cal G}_{ab} \bm{X}^{a} \bm{X}^{b} + \frac{i}{2\hbar}\tilde{{\cal G}}_{ab} (\bm{X}^{a} \bm{P}^b+\bm{P}^b \bm{X}^{a})
	-\frac{{\tilde{\tilde{\cal G}}}_{ab}}{\hbar^2\omega} \bm{P}^a \bm{P}^b\,,\\
	\label{QUOQ}
	\bm{{\cal Q}}_U^\omega &=& \frac{i}{\omega}\left(1-\frac{\omega^2}{2}(x^-)^2\right) {\cal G}_{--} -i x^- \omega  {\cal G}_{-a} \bm{X}^a+\frac{i x^- }{\hbar}\tilde{\cal G}_{-a} \bm{P}^{a}\,, \\
	\bm{{\cal Q}}_N^\omega&=&  x^- {\cal G}_{--} +    {\cal G}_{-a} \bm{X}^a - \frac{2}{\hbar\omega}\tilde{\cal G}_{-a} \bm{P}^{a} \,,
\end{eqnarray}
where $\bm{X}^a$ are operators. Notice that $\bm{{\cal O}}^\omega$, $\bm{{\cal Q}}_U^\omega$ and $\bm{{\cal Q}}_N^\omega$ are functions of $x^+$ because $\cal G$, $\tilde{\cal G}$ and $\tilde{\tilde{\cal G }}$ are function of $x^+$. The corrected state \eqref{correctedpsiHilbert} then follows from \eqref{PsiCompact2}:
\begin{equation}\label{Psi3O}
	\ket{\psi}= \sum_{s}\int d\omega  C^{s}(\omega) e^{-\frac{i E_{s} x^+}{\hbar}} e^{i\omega x^-} \left(1+\epsilon(\bm{{\cal O}}^\omega+ \bm{{\cal Q}}^\omega_U+\bm{{\cal Q}}^\omega_{N})\right)  \bm{a}^{\dagger}_{s}(\omega)\ket{0}+O(\epsilon^2)\,.
\end{equation}
Alice is at $x^+=0$. Notice that  $\bm{{\cal O}}^\omega$, $\bm{{\cal Q}}^\omega_U$ and $\bm{{\cal Q}}^\omega_N$ vanish at $x^+=0$ and we get,
\begin{equation}
	\ket{\psi_{\text{Alice}}}= \ket{\psi}|_{x^+=0}\,,
\end{equation}
Bob is at $x^+$, so the state he sees is given by,
\begin{equation}
	\ket{\psi_{\text{Bob}}}= \ket{\psi}|_{x^+=T}\,.
\end{equation}
Let us choose a normal distribution for $C^{s}(\omega)$ in terms of $\omega$ around $\omega_0$ with width of $\sigma$:
\begin{equation}
	C^{s}(\omega)= \frac{1}{\sigma\sqrt{2\pi}}e^{-\frac{(\omega-\omega_0)^2}{2\sigma^2}}\tilde{C}^{s}\,,
\end{equation}
where $\tilde{C}^{s}$ is a real number. We assume that $\sigma\ll \omega$. This allows us to simplify \eqref{Psi3O} to:
\begin{equation}\label{HilbertPsiAux}
	\ket{\psi}=(1+\epsilon\bm{{\cal O}}^{\omega_0}+\epsilon \bm{{\cal Q}}^{\omega_0}_U+\epsilon\bm{{\cal Q}}_{N}^{\omega_0}) e^{-\frac{i {\bm{H}} x^+}{\hbar}}\ket{\psi}|_{x^+=0}+O\left(\epsilon^2,\epsilon \frac{\sigma}{\omega}\right)\,.
\end{equation}
We notice that \eqref{HilbertPsiAux} can be utilized to write:
\begin{equation}\label{BobAliceHilbert}
	\ket{\psi_{\text{Bob}}}=(1+\epsilon\bm{{\cal O}}^{\omega_0}+\epsilon \bm{{\cal Q}}^{\omega_0}_U+\epsilon\bm{{\cal Q}}_{N}^{\omega_0}) e^{-\frac{i {\bm{H}} T}{\hbar}}\ket{\psi_{\text{Alice}}}+O\left(\epsilon^2,\epsilon \frac{\sigma}{\omega}\right)\,.
\end{equation}
We highlight that $1+\epsilon \bm{{\cal Q}}_{U}^{\omega_0}$ and $1+\epsilon \bm{{\cal O}}^{\omega_0}$  are unitary operators:
\begin{eqnarray}
	{\bm{{\cal O}}^{\omega_0}}^\dagger+\bm{{\cal O}}^{\omega_0} &=& 0  \,,\\
	{\bm{{\cal Q}}_U^{\omega_0}}^\dagger+\bm{{\cal Q}}_U^{\omega_0} &=& 0\,,
\end{eqnarray}
while $1+\epsilon\bm{{\cal Q}}_{N}^{\omega_0}$ is not unitary:
\begin{equation}
	{\bm{{\cal Q}}_N^{\omega_0}}^\dagger+\bm{{\cal Q}}_N^{\omega_0} \neq 0\,.
\end{equation} 
We refer to $\bm{{\cal Q}}_{U}$ and $\bm{{\cal O}}$ as the generators of quasi unitary operators, $\bm{{\cal Q}}_{N}$  represents the generator of a non-unitary. All of $\bm{{\cal Q}}_{U}$, $\bm{{\cal O}}$  and $\bm{{\cal Q}}_{N}$ are generators of non-local operators. 

Each transverse polarisation of the $U(1)$ gauge field gets corrected by the spacetime geometry as if it were a massless scalar. So \eqref{BobAliceHilbert} is valid for the  $U(1)$ guage field too.  This concludes the correction for the quantum realm. 
\section*{\textbf{\large{Supplementary Note 5: 	Propagation in the Schwarzschild spacetime geometry }}}\label{SI2}

In a spacetime endowed with the metric $g_{\mu\nu}$, a geodesic $x^{\mu}(\tau)$ can be obtained from an effective action: 
\begin{eqnarray}
	S&=& \int d\tau {\cal L}\,,\\
	{\cal L}&=& g_{\mu\nu} \dot{x}^\mu \dot{x}^\nu\,,
\end{eqnarray}
where $\tau$ is an affine parameter. For the Schwarzschild black hole in the standard coordinates, this is,
\begin{equation}
	{\cal L} = -\left(1-\frac{m}{r}\right)\dot{t}^2+ \frac{\dot{r}^2}{1-\frac{m}{r}}  + r^2 \left(\dot{\theta}^2+ \sin^2 \theta \dot{\phi}^2\right)\,,
\end{equation}
where $m ={2 G_N M_{\bullet}}$ that $M_{\bullet}$ is the mass of the black hole and $c=1$ is understood. We choose the units such that $m=1$. Due to the spherical symmetry, without loss of generality, we can choose the equatorial plane $\theta=\frac{\pi}{2}$ and $\dot{\theta}=0$ to describe any given geodesic at all times.  The cyclic variables of $\phi$ and $t$ lead to invariant quantities:
\begin{eqnarray}
	\frac{\partial {\cal L}}{\partial \phi} = 0 &\to& r^2 \dot{\phi} = L \,, \label{phidot}\\
	\frac{\partial {\cal L}}{\partial t} = 0 &\to& \left(1-\frac{1}{r}\right) \dot{t} = E\label{tdot}\,.
\end{eqnarray}
We consider null geodesics reaching the asymptotic infinity and set $E=1$.  Due to the form of the Lagrangian, its Legendre transformation, which is the Lagrangian itself, is invariant.  We consider a null geodesic, and set ${\cal L}=0$ giving,
\begin{equation}\label{rdot}
	|\dot{r}|= \gamma(r)= \sqrt{1-\frac{1}{r^2}\left(1-\frac{1}{r}\right) L^2}    \,.
\end{equation}
The non-zero components of the Riemann tensor on the geodesic ($\theta=\frac{\pi}{2}$) in the standard spherical coordinates  are:
\begin{subequations}\label{RiemannSchStandardCoordinated}
	\begin{eqnarray}
		R_{trtr}&=&-\frac{1}{r^3}\,,\\
		R_{\theta\phi\theta\phi}&=& r\,\\
		R_{t\theta t\theta} & =& R_{t\phi t\phi} = \frac{r-1}{2r^2}\,,\\
		R_{r\theta r\theta} &=&R_{r\phi r\phi} =  -\frac{1}{2\left(r-1\right)}\,.
	\end{eqnarray}
\end{subequations}
The coordinate independent representation of the Riemann tensor, therefore, follows: 
\begin{eqnarray}\label{RC}
	R=&-&\frac{1}{r^3} \left(dt\!\wedge\!dr\right)\otimes \left(dt\!\wedge\!dr\right)+ r\left(d\theta\!\wedge\!d\phi\right)\otimes\left(d\theta\!\wedge\!d\phi\right)+\frac{r-1}{2r^2} \left((dt\!\wedge\!d\phi)\otimes\left(dt\!\wedge\!d\phi\right)+ \left(dt\!\wedge\!d\theta\right)\otimes\left(dt\!\wedge\!d\theta\right)\right) \nonumber\\
	&-&\frac{1}{2(r-1)} \left((dr\!\wedge\!d\phi)\otimes(dr\!\wedge\!d\phi)+ (dr\!\wedge\!d\theta)\otimes(dr\!\wedge\!d\theta)\right)\,.
\end{eqnarray}
We would like to compute the components of the Riemann tensor in $x^+,x^{\bar{a}}$ coordinates. To this aim, we first consider an infinitesimal displacement $\delta\vec{x}$:
\begin{eqnarray}\label{dxvs}
	d\vec{x} = f dt  \hat{\bm{e}}_t + \frac{dr}{f}\hat{\bm{e}}_r + r d\phi \hat{\bm{e}}_\phi + r d\theta \hat{\bm{e}}_\theta \,,
\end{eqnarray}
where $\hat{\bm{e}}_t$, $\hat{\bm{e}}_r$, $\hat{\bm{e}}_\phi$ and $\hat{\bm{e}}_\theta$ are the normalized unit vectors in $t,r,\theta,\phi$ coordinates, and,
\begin{equation}\label{fr}
	f= \sqrt{1-\frac{1}{r}}\,.
\end{equation} 
The displacement vector in the $x^+,x^{\bar{a}}$ coordinates follows
\begin{eqnarray}\label{dxvx}
	d\vec{x} =  dx^+ \hat{\bm{e}}_+ + dx^- \hat{\bm{e}}_- + dx^1 \hat{\bm{e}}_1 + dx^2 \hat{\bm{e}}_2 \,,
\end{eqnarray}
where $\hat{\bm{e}}_+$, $\hat{\bm{e}}_-$, $\hat{\bm{e}}_1$ and $\hat{\bm{e}}_2$ are the normalized unit vectors in the $x^+,x^{\bar{a}}$ coordinates. We are not interested in explicit reproduction of the gravitational redshift. So instead of setting $\hat{\bm{e}}_+$  equal to the the tangent of the geodesic, we set $\hat{\bm{e}}_+$ proportional to the tangent of the geodesic where the vectors of $\hat{\bm{e}}_-$, $\hat{\bm{e}}_+$, $\hat{\bm{e}}_1$ and $\hat{\bm{e}}_2$ are:
\begin{subequations}\label{e_x_r2}
	\begin{eqnarray}
		\hat{\bm{e}}_+ &=& \frac{1}{\sqrt{2}}\left(+\hat{\bm{e}}_t+   \dot{r}\hat{\bm{e}}_r + rf \dot{\phi}\hat{\bm{e}}_\phi\right)\,,  \\
		\hat{\bm{e}}_- &=&\frac{1}{\sqrt{2}}\left(-\hat{\bm{e}}_t+  \dot{r} \hat{\bm{e}}_r + r f \dot{\phi} \hat{\bm{e}}_\phi\right)\,, \\
		\hat{\bm{e}}_1 &=& - r f \dot{\phi} \hat{\bm{e}}_r +\dot{r} \hat{\bm{e}}_\phi \,, \\
		\hat{\bm{e}}_2 &=& \hat{\bm{e}}_\theta\,,
	\end{eqnarray}
\end{subequations}
We use \eqref{e_x_r2} in \eqref{dxvs} and equate it to \eqref{dxvx}, and obtain:
\begin{subequations}\label{dxGeneral}
	\begin{eqnarray}
		d t &=&\frac{1}{f\sqrt{2}}\left(d x^+-d x^-\right)\,,\\
		d r &=& \frac{f\dot{r}}{\sqrt{2}} \left(d x^++d x^-\right)- r f^2 \dot{\phi} d x^1\,, \\
		d\phi & = &  \frac{f \dot{\phi}}{\sqrt{2}} \left(d x^+ +d x^-\right)+ \frac{\dot{r}}{r} d x^1\,,\\
		d\theta &=& \frac{1}{r}d x^2\,.
	\end{eqnarray}
\end{subequations}
We would like to compute $R_{+\bar{a}+\bar{b}}$. In doing so, we keep only the linear $dx^+$ terms in the wedge products that appeared in \eqref{RC}:
\begin{subequations}\label{partialWedge}
	\begin{eqnarray}
		dt\wedge dr &=& \dot{r} dx^+\wedge dx^- - \frac{rf\dot{\phi}}{\sqrt{2}} dx^+ \wedge dx^1\,,\\
		dt\wedge d\phi &=&  \dot{\phi} dx^+\wedge dx^-+ \frac{\dot{r}}{f r \sqrt{2}} dx^+\wedge dx^1\,,\\
		dt\wedge d\theta &=& \frac{1}{r f\sqrt{2}}d x^+\wedge d x^2\,,\\
		dr\wedge d\phi &=&  \frac{f}{r\sqrt{2}} dx^+\wedge dx^1\,,\\
		dr\wedge d\theta &=& \frac{f\dot{r}}{r\sqrt{2}} dx^+\wedge dx^2\,,\\
		d\theta\wedge d\phi &=&- \frac{f\dot{\phi}}{r\sqrt{2}} dx^+ \wedge dx^2\,.
	\end{eqnarray}
\end{subequations}
Utilizing \eqref{partialWedge} in \eqref{RC} enables us to extract the components of $R_{+\bar{a}+\bar{b}}$:
\begin{subequations}\label{RPaPb}
	\begin{eqnarray}
		R_{+-+-}&=& -\frac{\dot{r}^2}{r^3} + \frac{r-1}{2r^2}\dot{\phi}^2\,, \\
		R_{+1+1} &=&-\frac{1}{4r^3}+\frac{\dot{r}^2}{4r^3}- \frac{f^2 \dot{\phi}^2}{2 r}\,,\\
		R_{+2+2} &=&\frac{1}{4r^3}-\frac{\dot{r}^2}{4r^3}+ \frac{f^2 \dot{\phi}^2}{2 r}\,,\\
		R_{+-+1} &=&\textcolor{black}{ \frac{3\sqrt{2}}{2r^2}  f \dot{r} \dot{\phi}\,,}\\
		R_{+-+2}&=& R_{+1+2}=0\,.
	\end{eqnarray}
\end{subequations}
As a consistency check, we observe that \eqref{RPaPb} satisfies \eqref{RicciPlusPlus}, and $R_{+-+2}$ and $R_{+1+2}$ vanish due to parity symmetry in the $x^2$ direction: $x^2\to -x^2$. We use \eqref{phidot}, \eqref{rdot} and \eqref{fr} to simplify \eqref{RPaPb}:
\begin{subequations}\label{RPaPbSimplified}
	\begin{eqnarray}\label{Rpmpm}
		R_{+-+-}&=& \frac{3 L^2(r-1)}{2r^6}-\frac{1}{r^3}\,, \\
		R_{+1+1} &=&-\frac{3L^2(r-1)}{4r^6} \,,\\
		R_{+2+2} &=&\frac{3L^2(r-1)}{4r^6} \,,\\
		R_{+-+1} &=&\textcolor{black}{  \frac{3L}{2r^4} \sqrt{2-\frac{2}{r}}  \dot{r}}\,,\\
		R_{+-+2}&=& R_{+1+2}= 0\,.
	\end{eqnarray}
\end{subequations}
Eq. \eqref{RPaPbSimplified} enables us to compute the $\cal G$ coefficients given \eqref{eq21aux} for any geodesic.

\subsection{Propagating along a general null geodesic}
\label{PropgationGeneralGeodesic}
\begin{figure}[t]
	\centering
	\includegraphics[width=10cm]{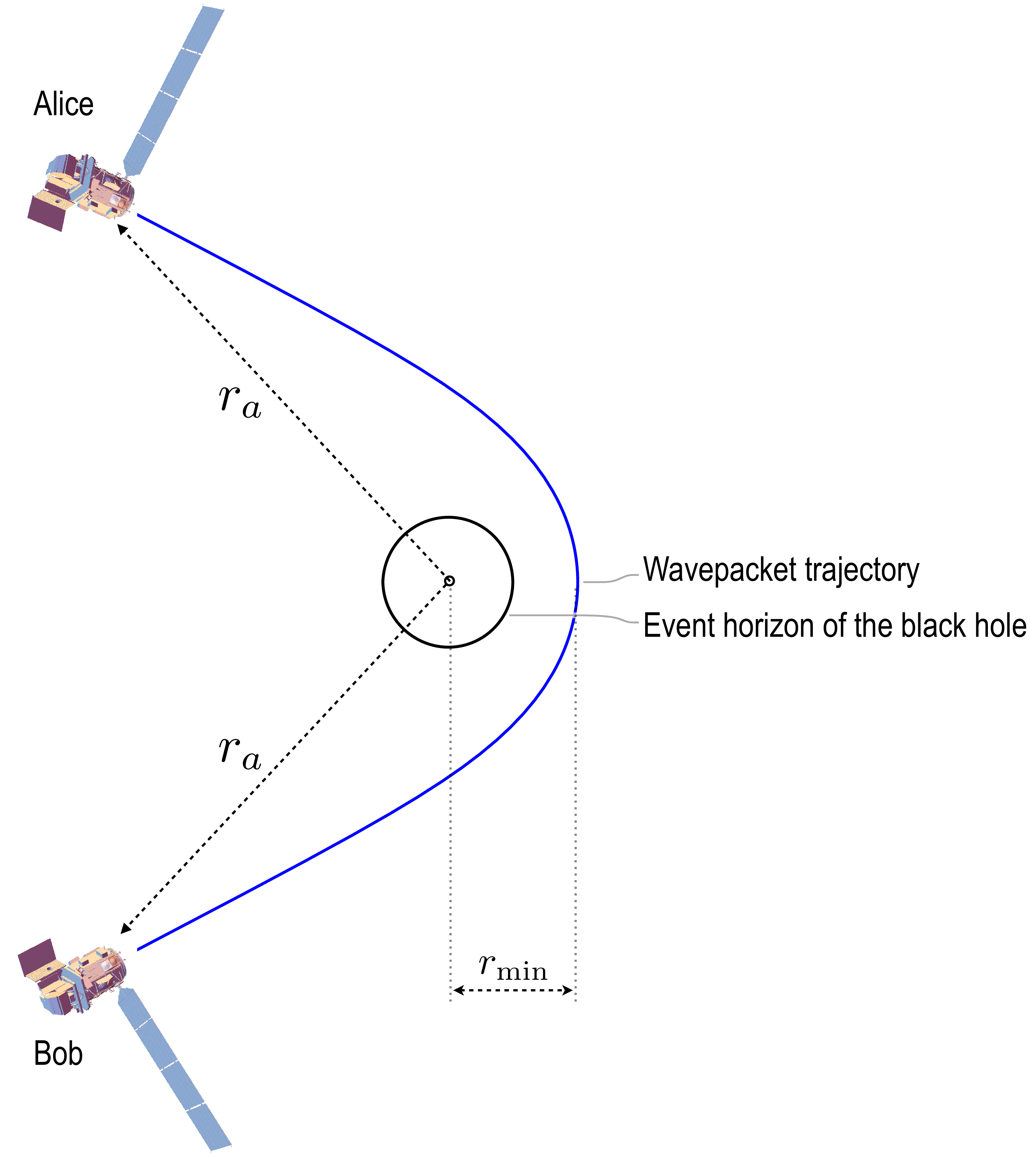}
	\caption{The geometry of communication for an arbitrary geodesic path, where two satellites share a wavepacket. The wavepacket traverses through the spacetime, goes near the black hole, and is deflected toward Bob.}
	\label{fig:GG}
\end{figure}
This section considers that Alice and Bob are at the same far distance $r_a$ from the black hole. Alice sends a wavepacket toward Bob.  The wavepacket goes near the black hole and is deflected toward Bob, fig. \ref{fig:GG}.  The package takes $x^+= T$ of the affine parameter  to reach Bob. We notice that $\dot{r}$ is negative for $x^+\in\{0,\frac{T}{2}\}$ and is positive for $x^+\in\{\frac{T}{2},T\}$.  To calculate the integration of the $\tilde{R}_{+\bar{a}+\bar{b}}$ over the geodesic, we divide integration to intervals of positive and negative $\dot{r}$ :
\begin{eqnarray}
	{\cal G}_{\bar{a}\bar{b}} &=& 
	\textcolor{red}{\int_0^{\frac{T}{2}} d\tau R_{+\bar{a}+\bar{b}}}
	+\textcolor{blue}{ \int^T_{\frac{T}{2}} d\tau R_{+\bar{a}+\bar{b}}}  =\textcolor{red}{\int_{r_a}^{r_{\text{min}}} \frac{dr}{\dot{r}} R_{+\bar{a}+\bar{b}}}
	+\textcolor{blue}{ \int^{r_a}_{r_{\text{min}}} \frac{dr}{\dot{r}} R_{+\bar{a}+\bar{b}}}+\textcolor{red}{\int^{r_a}_{r_{\text{min}}} \frac{dr}{\gamma(r)} R_{+\bar{a}+\bar{b}}}
	+ \textcolor{blue}{\int^{r_a}_{r_{\text{min}}} \frac{dr}{\gamma(r)} R_{+\bar{a}+\bar{b}}}\,,
\end{eqnarray}
where red color is used for the interval where $\dot{r}$ is negative, and  blue is used when $\dot{r}$ is positive. In the last line, the integration limit of integrals are written from a smaller value of $r$ to a larger one, and $\dot{r}=\gamma(r)$ is used for blue integral, and $\dot{r}=-\gamma(r)$ is used for red integral where $\gamma$ is defined in \eqref{rdot}. Utilizing \eqref{RPaPbSimplified} then gives
\begin{subequations}
	\label{G}
	\begin{eqnarray}
		{\cal G}_{--} &=&2 \int_{r_\text{min}}^{r_a} \frac{dr}{\gamma(r)} R_{+-+-},\\
		{\cal G}_{-1} &=& 0 \,,\\
		{\cal G}_{11} &=& - {\cal G}_{22}=2 \int_{r_\text{min}}^{r_a} \frac{dr}{\gamma(r)} R_{+1+1}  \,.
		\label{G11}
	\end{eqnarray}
\end{subequations}
where $\gamma(r)$ is defined in \eqref{rdot}.
We also need to calculate $\tilde{\cal G}_{\bar{a}\bar{b}}$ in \eqref{eq21aux}:
\begin{eqnarray}
	\tilde{\cal G}_{\bar{a}\bar{b}} = \int_0^T d\tau \int_0^\tau d\tau' R_{+\bar{a}+\bar{b}}(\tau') \,,
\end{eqnarray}
where $T$ is the total amount of affine parameter that the light packet takes to reach from Alice to Bob. We would like to change the integration over $\tau$ to an integration over $r$ by $d\tau=\frac{dr}{\dot{r}}$. We break the integral to intervals of constant sign for $\dot{r}$,
\begin{eqnarray}
	\tilde{\cal G}_{\bar{a}\bar{b}} &=& 
	\textcolor{red}{\int_0^\frac{T}{2} d\tau \int_0^\tau d\tau' R_{+\bar{a}+\bar{b}}} 
	+ \textcolor{blue}{\int_\frac{T}{2}^T d\tau}\textcolor{red}{\int_0^{\frac{T}{2}} d\tau'  R_{+\bar{a}+\bar{b}}}
	+\textcolor{blue}{\int_\frac{T}{2}^T d\tau\int_{\frac{T}{2}}^\tau d\tau' R_{+\bar{a}+\bar{b}}} \,,
\end{eqnarray}
where red color is for the interval of negative $\dot{r}$ and blue colour is for positive $\dot{r}$. Next we change the integral from $\tau$ to $r$ through $d\tau=\frac{dr}{\dot{r}}$:
\begin{eqnarray}
	\tilde{\cal G}_{\bar{a}\bar{b}}&=&
	\textcolor{red}{\int_{r_a}^{r_{\text{min}}} \frac{dr}{\gamma(r)}\int_{r_a}^{r} \frac{dr'}{\gamma(r')}R_{+\bar{a}+\bar{b}}} 
	-\textcolor{blue}{\int^{r_a}_{r_{\text{min}}} \frac{dr}{\gamma(r)}}
	\textcolor{red}{\int^{r_\text{min}}_{r_a}\frac{dr'}{\gamma(r')}R_{+\bar{a}+\bar{b}}}+\textcolor{blue}{\int^{r_a}_{r_{\text{min}}} \frac{dr}{\gamma(r)}\int_{r_\text{min}}^{r}\frac{dr'}{\gamma(r')}R_{+\bar{a}+\bar{b}}}\,,
\end{eqnarray}
where  $\dot{r}=\gamma(r)$ is used for the blue integral,   $\dot{r}=-\gamma(r)$ is used for red integrals, and $\gamma$ is defined in \eqref{rdot}. We write all the integrals from a smaller value of radius to the larger one,
\begin{eqnarray}
	\label{GTildeGeneralGeodesic}
	\tilde{\cal G}_{\bar{a}\bar{b}}&=&
	\textcolor{red}{\int^{r_a}_{r_{\text{min}}} \frac{dr}{\gamma(r)}\int^{r_a}_{r} \frac{dr'}{\gamma(r')}R_{+\bar{a}+\bar{b}}} 
	+\textcolor{blue}{\int^{r_a}_{r_{\text{min}}} \frac{dr}{\gamma(r)}}
	\textcolor{red}{\int_{r_\text{min}}^{r_a}\frac{dr'}{\gamma(r')}R_{+\bar{a}+\bar{b}}}+\textcolor{blue}{\int^{r_a}_{r_{\text{min}}} \frac{dr}{\gamma(r)}\int_{r_\text{min}}^{r}\frac{dr'}{\gamma(r')}R_{+\bar{a}+\bar{b}}}\,.
\end{eqnarray}
Utilizing \eqref{RPaPbSimplified}  then gives:
\begin{subequations}
	\label{GT}
	\begin{eqnarray}
		\tilde{\cal G}_{-1} &=& -2\int_{r_{\text{min}}}^{r_a} \frac{dr}{\gamma(r)} \int_{r}^{r_a} dr' \frac{R_{+-+1}}{\dot{r}}  \,,
		\\
		\label{GT11}
		\tilde{\cal G}_{11} &=&-\tilde{\cal G}_{22} = \left(\int^{r_a}_{r_{\text{min}}} \frac{dr}{\gamma(r)}\right){\cal G}_{11}\,.
	\end{eqnarray}
\end{subequations}
To compute  ${\tilde{{\tilde{\cal G}}}}_{\bar{a}\bar{b}}$, notice  that it is given by 
\begin{equation}
	\tilde{\tilde{{\cal G}}}_{\bar{a}\bar{b}} = \int_{0}^T d\tau_1  \int_0^{\tau_1} d\tau_2\int_{0}^{\tau_2} d\tau_3 \tilde{R}_{\bar{a}\bar{b}}(\tau_3)\,.
\end{equation}
We divide the domain of integration to intervals that have constant sign of $\dot{r}$,
\begin{eqnarray}
	\tilde{\tilde{{\cal G}}}_{\bar{a}\bar{b}} &=& 
	\textcolor{red}{\int_{0}^{\frac{T}{2}}\!\!d\tau_1  \int_0^{\tau_1}\!\! d\tau_2\int_{0}^{\tau_2} \!\!d\tau_3 R_{+\bar{a}+\bar{b}}(\tau_3)}
	+  \textcolor{blue}{\int_{\frac{T}{2}}^T\!\! d\tau_1}  \textcolor{red}{\int_0^{\frac{T}{2}}\!\! d\tau_2\int_{0}^{\tau_2}\!\! d\tau_3 R_{+\bar{a}+\bar{b}}(\tau_3)}
	\nonumber\\
	&+&\textcolor{blue}{\int_{\frac{T}{2}}^T\!\! d\tau_1  \int_{\frac{T}{2}}^{\tau_1}\!\! d\tau_2}\textcolor{red}{\int_{0}^{\frac{T}{2}} \!\!d\tau_3 R_{+\bar{a}+\bar{b}}(\tau_3)}
	+\textcolor{blue}{\int_{\frac{T}{2}}^T\!\! d\tau_1  \int_{\frac{T}{2}}^{\tau_1}\!\! d\tau_2\int_{\frac{T}{2}}^{\tau_2}\!\!  d\tau_3 R_{+\bar{a}+\bar{b}}(\tau_3)}\,,
\end{eqnarray}
where the red color is associated to intervals where $\dot{r}$ is negative, blue is associated to the intervals wherein $\dot{r}$ is positive. Now we change the integral from over $\tau$ to $r$: 
\begin{eqnarray}
	\tilde{\tilde{{\cal G}}}_{\bar{a}\bar{b}} &=& 
	-\textcolor{red}{\int_{r_a}^{r_{\text{min}}}\!\frac{dr_1}{\gamma(r_1)}  \int_{r_a}^{r_1} \!\frac{dr_2}{\gamma({r}_2)} \int_{r_a}^{r_2} \frac{dr_3}{\gamma({r}_3)} R_{+\bar{a}+\bar{b}}}+ \textcolor{blue}{\int_{r_{\text{min}}}^{r_a}\! \frac{dr_1}{\gamma(r_1)}}  \textcolor{red}{\int_{r_a}^{r_{\text{min}}}\! \frac{dr_2}{\gamma(r_2)} 
		\int_{r_a}^{r_2} \frac{dr_3}{\gamma({r}_3)} R_{+\bar{a}+\bar{b}}}\nonumber\\
	&-&\textcolor{blue}{\int_{r_\text{min}}^{r_a}\! \frac{dr_1}{\gamma({r}_1)}  \int_{r_\text{min}}^{r_1}\! \frac{dr_2}{\gamma({r}_2)}}\textcolor{red}{\int_{r_a}^{r_\text{min}}\! \frac{dr_3}{\gamma({r}_3)} R_{+\bar{a}+\bar{b}}}+\textcolor{blue}{\int^{r_a}_{r_\text{min}}\! \frac{dr_1}{\gamma({r}_1)}  \int_{r_{\text{min}}}^{r_1}\! \frac{dr_2}{\gamma({r}_2)}\int_{r_\text{min}}^{r_2}\! \frac{dr_3}{\gamma({r}_3)} R_{+\bar{a}+\bar{b}}}\,,
\end{eqnarray}
where  $\dot{r}=\gamma(r)$ is used for the blue integrals,   $\dot{r}=-\gamma(r)$ is used for red integrals, and $\gamma$ is defined in \eqref{rdot}.  We interchange the domain of all the integrals, such that all the integrals start from the smaller value of $r$:
\begin{eqnarray}
	\tilde{\tilde{{\cal G}}}_{\bar{a}\bar{b}} &=& 
	\textcolor{red}{\int^{r_a}_{r_{\text{min}}}\!\frac{dr_1}{\gamma(r_1)}  \int^{r_a}_{r_1} \!\frac{dr_2}{\gamma({r}_2)} \int^{r_a}_{r_2} \frac{dr_3}{\gamma({r}_3)} R_{+\bar{a}+\bar{b}}}+ \textcolor{blue}{\int_{r_{\text{min}}}^{r_a}\! \frac{dr_1}{\gamma(r_1)} }  \textcolor{red}{\int^{r_a}_{r_{\text{min}}}\! \frac{dr_2}{\gamma(r_2)} 
		\int^{r_a}_{r_2} \frac{dr_3}{\gamma({r}_3)} R_{+\bar{a}+\bar{b}}}\nonumber\\
	&+&\textcolor{blue}{\int_{r_\text{min}}^{r_a}\! \frac{dr_1}{\gamma({r}_1)}  \int_{r_\text{min}}^{r_1}\! \frac{dr_2}{\gamma({r}_2)}}\textcolor{red}{\int^{r_a}_{r_\text{min}}\! \frac{dr_3}{\gamma({r}_3)} R_{+\bar{a}+\bar{b}}}+\textcolor{blue}{\int^{r_a}_{r_\text{min}}\! \frac{dr_1}{\gamma({r}_1)}  \int_{r_{\text{min}}}^{r_1}\! \frac{dr_2}{\gamma({r}_2)}\int_{r_\text{min}}^{r_2}\! \frac{dr_3}{\gamma({r}_3)} R_{+\bar{a}+\bar{b}}}\,.
\end{eqnarray}
Noticing that $R_{+1+1}$ does not depend on $\dot{r}$, and after simplification we then obtain: 
\begin{eqnarray}
	\label{GTT}
	\tilde{\tilde{{\cal G}}}_{11} = -\tilde{\tilde{{\cal G}}}_{2 2} &=& \left(\int_{r_{\text{mim}}}^{r_a} \!\!\frac{dr}{\gamma(r)}\right)^2 {\cal G}_{11}-\left(\int_{r_{\text{min}}}^{r_a} \frac{dr_1}{\gamma(r_1)}\int_{r_1}^{r_a}\frac{dr_2}{\gamma(r_2)}\right) {\cal G}_{11} \nonumber\\
	&+&2
	\int^{r_a}_{r_{\text{min}}}\!\! \frac{dr_1}{\gamma({r}_1)} 
	\int^{r_a}_{r_1}\!\! \frac{dr_2}{\gamma({r}_2)}
	\int^{r_a}_{r_2}\!\! \frac{dr_3}{\gamma({r}_3)} R_{+1+1}(r_3)\,.
\end{eqnarray}
Notice that in the simplification we haven't tried to get the minimum number of integrations. We have ordered the boundary of all integrals from a lower value of $r$ to $r_a$. Eq. \eqref{RPaPbSimplified}, \eqref{G}, \eqref{GT} and \eqref{GTT} identify all the $\cal G$-coefficients of $\cal O$ \eqref{OO}, ${\cal Q}_U$ \eqref{OQU} and ${\cal Q}_N$ \eqref{OQN} at Bob's position for a general geodesic.

\section*{\textbf{\large{Supplementary Note 6: Deflection by the Sun}}}\label{SI3}
\begin{figure}
	\centering
	\includegraphics[width=\textwidth]{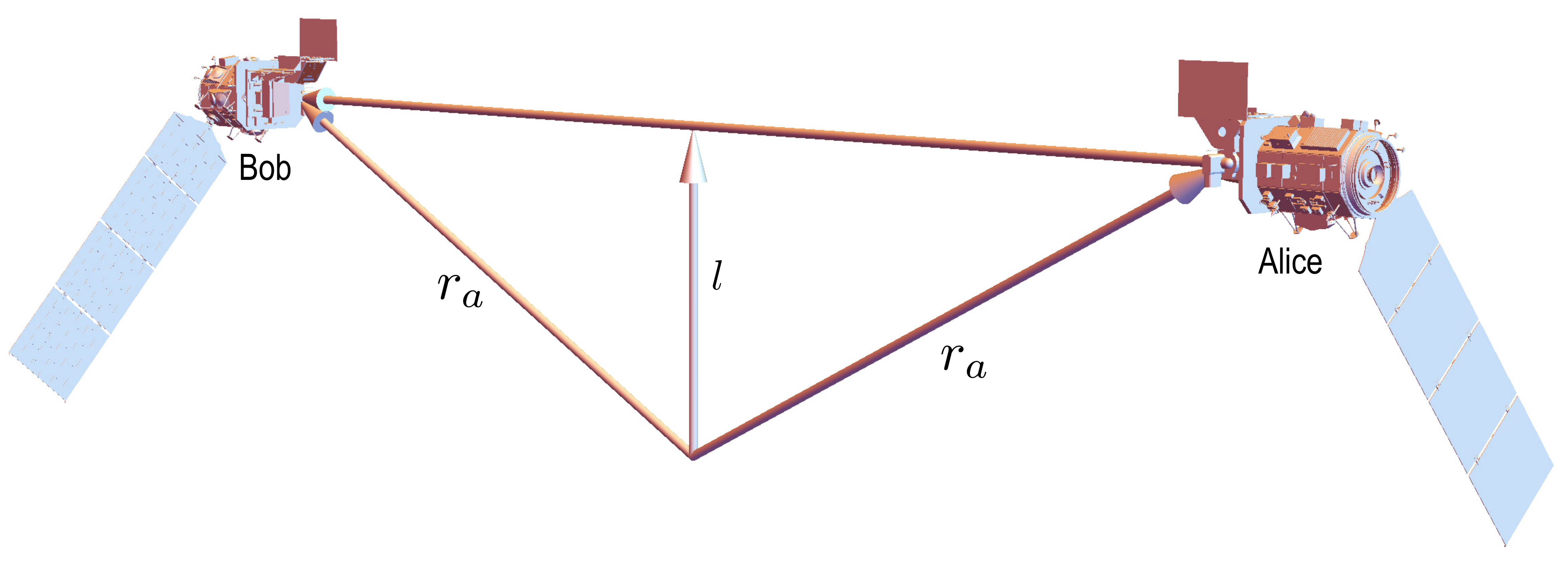}
	\caption{Alice and Bob are in the weak gravity regime at distance of $r_a$ from the central mass. Alice sends a pulse beam (contains information) toward Bob. The wavepacket reaches distance of $l$ to the central mass. $l$ is large and the wavepackage is very slightly deflected. Its trajectory remains almost a straight line.}
	\label{fig:WeakGravity}
\end{figure}

This section considers the communication between Alice and Bob in the weak regime of gravity of the Schwarzschild spacetime geometry, see Fig.~\ref{fig:WeakGravity}. 
Alice sends a wavepacket/beam toward Bob. The wavepacket reaches a distance of $\textcolor{black}{r_{min}=}l$ from the central mass and $l$ is large. So the packet is very slightly deflected. Its trajectory remains almost a straight line. In so doing, we treat the Schwarzschild radius as a small parameter, and \textcolor{black}{approximate $\gamma$ defined in \eqref{rdot} to
	\begin{equation}
		\gamma= \sqrt{1-\frac{l^2}{r^2}},
	\end{equation} 
	and} simplify  \eqref{G}, \eqref{GT} and \eqref{GTT} to,
\begin{subequations}\label{GfactorWeakGravity}
	\begin{eqnarray}
		l^{2}{\cal G}_{--} &=&\frac{\sqrt{a^2-1}}{a^3}\,,\\
		l^2{\cal G}_{-1} &=&0\,,\\
		\label{G11_large_a}
		l^{2}{\cal G}_{11} &=&-l^{2}{\cal G}_{22}=\frac{\sqrt{a^2-1}}{2a^3}(1+2a^2) \,,\\
		l\tilde{\cal G}_{-1} &=&\textcolor{black}{-\frac{(a-1)\sqrt{2(a^2-1)}}{2 a}}\,,\\
		l\tilde{\cal G}_{11} &=&-l\tilde{\cal G}_{22}=\frac{1}{2a^3}(a^2-1)(1+2a^2)\,,\\
		{\tilde{\tilde{\cal G}}}_{11} &=& -{\tilde{\tilde{\cal G}}}_{22}=-\frac{(a^2-1)^{\frac{3}{2}}(1+a^2)}{2a^3}\,,
	\end{eqnarray}
\end{subequations}
where $a={r_a}/{l}$ is a dimensionless parameter $a$. Utilizing \eqref{GfactorWeakGravity} in \eqref{OO}, \eqref{OQU} and \eqref{OQN} yields:
\begin{subequations}\label{OsWeak}
	\begin{eqnarray}
		{\cal O}^\omega &=&- \frac{i\omega}{2}  {\cal G}_{11} (x_1^2-x_2^2)+ \tilde{{\cal G}}_{11} (x_{1} \partial^1-x_2\partial^2)  
		+\frac{i}{\omega} \tilde{\tilde{{\cal G}}} _{11}\left((\partial^1)^2- (\partial^2)^2\right)\\
		{\cal Q}_U^\omega &=&\frac{i}{\omega}\left(1-\frac{\omega^2}{2}(x^-)^2\right) {\cal G}_{--} +x^- \tilde{\cal G}_{-1} {\partial}^{1}\,,\\
		{\cal Q}_N^{\omega}&=&  x^- {\cal G}_{--}  + \frac{2i}{\omega}\tilde{\cal G}_{-1} \partial^{1}\,.
	\end{eqnarray}
\end{subequations}
Employing them in \eqref{PsiCompact2} results in the corrected wave:
\begin{eqnarray}\label{PsiGm1WeakGravity}
	\psi(x^\mu)&=&\int d\omega~ e^{i\omega x^-} \left(1+\epsilon {\cal O}^\omega \textcolor{black}{+}\epsilon{\cal Q}_U^\omega + \epsilon{\cal Q}_N^\omega\right)f^{(0)}_\omega+O(\epsilon^2)\,.
\end{eqnarray}
We assume that Alice prepares a frequency Gaussian wavepacket with the central frequency $\omega_0$ and  width $\sigma$:
\begin{equation}
	f^{(0)}_\omega= \frac{1}{\sigma \sqrt{2\pi}} e^{-\frac{(\omega-\omega_0)^2}{2\sigma^2}} f^{(0)}(x^+,x^1,x^2)\,,
\end{equation}
where $f^{(0)}$ satisfies the wave equation and $\sigma\ll\omega_0$. This assumption allows us to replace $\omega$ in the parentheses of the integral of \eqref{PsiGm1WeakGravity} with ${\omega}_0$, and perform the integration over $\omega$ to obtain:
\begin{eqnarray}\label{PsiGm1WeakGravity2}
	\psi(x^\mu)&=&e^{i \omega_0x^-} e^{-\frac{(\sigma x^-)^2 }{2}}\left(1+\epsilon \left({\cal O}^{\omega_0} + {\cal Q}_U^{\omega_0} + {\cal Q}_N^{{\omega}_0}\right)\right)f^{(0)}+O\left(\epsilon^2,  \frac{\epsilon\sigma}{{\omega}_0}\right) \,.
\end{eqnarray}
Alice sends a beam with a Hermite-Gaussian distribution in the transverse directions of $x^1$ and $x^2$, note that $f^{(0)}_\omega(x^+,x^a):=f_{\omega,m,n,q}^{(0)}=\text{HG}_{q}(x^+)\,\text{HG}_{m,n}(x_1,x_2)$. To represent such a solution, let $w_0$ be the initial width of the beam, $z_R$ be the Rayleigh range, and $w(x^+)$ be the width of the packet at $x^+$
\begin{eqnarray}
	\label{wz}
	w(x^+)&=& w_0 \sqrt{1+ \left(\frac{x^+}{z_R}\right)^2}\,,\\
	z_R &=& \frac{1}{2}{\omega}_0 w_0^2\,.
\end{eqnarray}
Define $\delta(x^+)$ as the Gouy phase:
\begin{equation}
	\delta(x^+)=\left(m+n+1\right) \, \arctan \left(\frac{x^+}{z_\mathrm{R}} \right)\,.
\end{equation}
For Hermite-Gaussian modes of $m,n$ around the mean value of $\omega_0$ with the width of $\sigma$:
\begin{subequations}\label{fzxy}
	\begin{eqnarray}
		\psi_{\text{Alice}}(x^\mu)&=&C_{m,n}\, e^{i {\omega}_0 x^-} e^{-\frac{\left(\sigma x^-\right)^2 }{2}} \left(\frac{w_0}{w(x^+)}\right) \text{H}_m\left(\frac{\sqrt{2}x_1}{w(x^+)}\right)\, \text{H}_n\left(\frac{\sqrt{2}x_2}{w(x^+)}\right)\,\exp\!\left(\!-\frac{{\omega}_0(x_1^2+x_2^2)}{2 z_R-2 i x^+}+i\delta(x^+)\!\right),
	\end{eqnarray}
\end{subequations}
where $\text{H}_n(.)$ are Hermite polynomials, and $m,n$ are positive integer values that define the transverse modes. Figure~4-({\bf a}) shows the intensity profile of Hermite-Gauss modes for first 9 modes, $m,n\in\{0,1,2\}$. Note that \eqref{fzxy} represents exact solutions for beams propagating in Minkowski spacetime geometry. We express \eqref{OsWeak} at $x^+=T$ in terms of $x$ and $y$:
\begin{subequations}\label{OsWeak1}
	\begin{eqnarray}
		{\cal O} &=&- i\omega_0 {w^2(T)}  {\cal G}_{11} (x^2-y^2) + \tilde{{\cal G}}_{11} (x \partial_x-y\partial_y) +\frac{2i  \tilde{\tilde{{\cal G}}}_{11}}{\omega_0 w^2(T)}\big(\partial_x^2- \partial^2_y\big)\,,\\
		{\cal Q}_U &=&\frac{i}{\omega_0}(1-\frac{\omega_0^2}{2}(x^-)^2){\cal G}_{--} +\frac{\sqrt{2}\tilde{\cal G}_{-1}}{w(T)}x^-  {\partial}_x\,,\\
		{\cal Q}_N&=&  x^- {\cal G}_{--}  + \frac{2\sqrt{2}i}{\omega_0 w(T)} \tilde{\cal G}_{-1}\partial_x\,,
	\end{eqnarray}
\end{subequations}
where $x={\sqrt{2} x_1}/{w(x^+)}$ and $y={\sqrt{2} x_2}/{w(x^+)}$ are dimensionless coordinates, and 
\textcolor{black}{
	\begin{eqnarray}
		T= 2 l \sqrt{a^2-1},
	\end{eqnarray}
}
is the time at which the wavepacket reaches to Bob -- for simplicity, we have dropped the superscript of $\omega_0$. Thus, the wavepacket at $T$ is,
\begin{equation}
	\psi_{\text{Bob}}(x^\mu)=\left(\left(1+ \epsilon\left({\cal O}+ {\cal Q}_U+ {\cal Q}_N\right)\right) \psi_{\text{Alice}}\right)|_{x^+=T}+ O\left(\epsilon^2,\epsilon \sigma\right)\,.
\end{equation}
\begin{figure}
	\centering
	\includegraphics[width=0.7\textwidth]{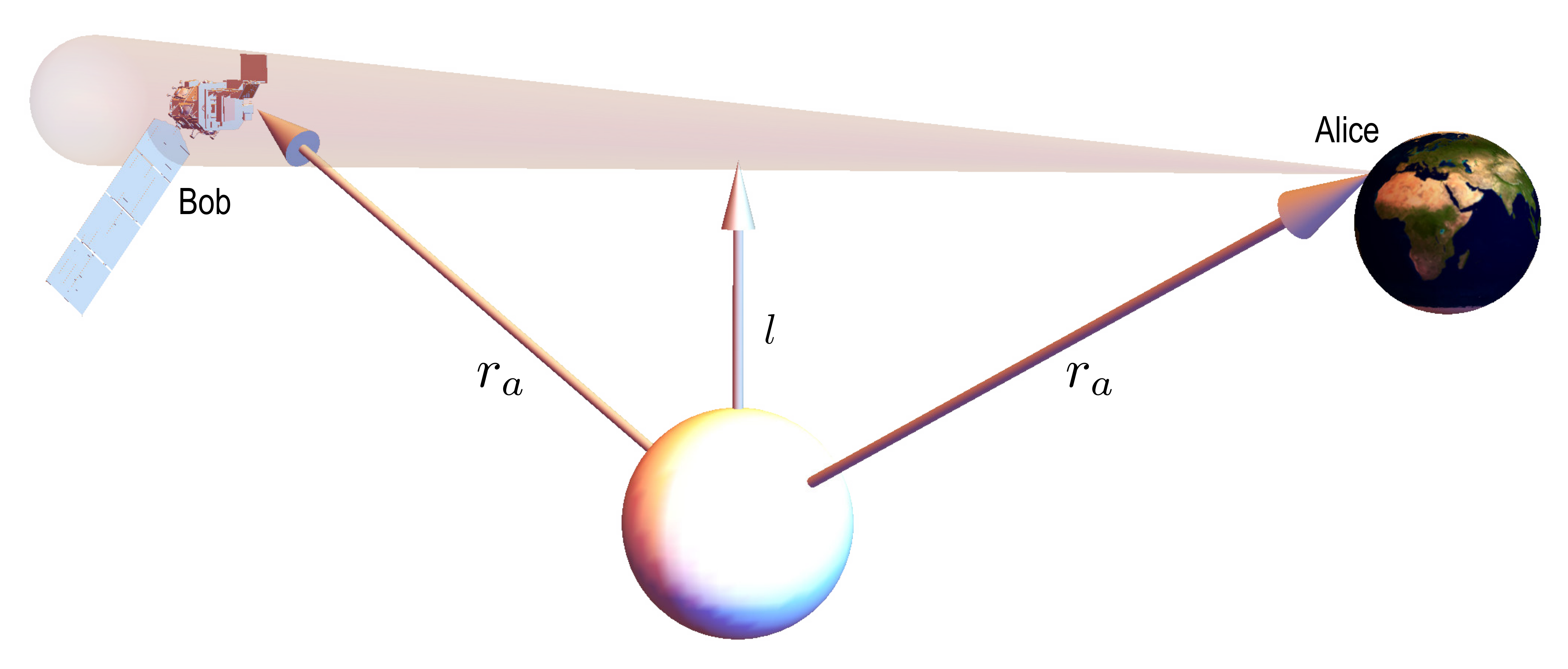}
	\caption{Alice uses the large telescopes on Earth, and sends a Hermite Gaussian signal toward Bob such that the width of the signal remains smaller than the Sun's radius at its closest encounter to the Sun. } 
	\label{fig:AliceBobNearSun}
\end{figure}
We would like to study the correction for the solar system, when Alice and Bob are at the mean-Earth-Sun distance from the sun and the wavepacket passes at $l=2R_\odot$, Fig.~\ref{fig:AliceBobNearSun}. We assume that $z_R\ll T$. So we set:
\begin{eqnarray}
	m & =& m_\odot= 2953\, \text{meters}\,,\\
	r_a&=& 1.51 \times 10^{11}\, \text{meters}\,,\\
	l &=& 2 R_\odot = 1.39\times 10^9 \text{meters}\,,\\
	a &=& 218.6\,.
\end{eqnarray}
This allows us to approximate:
\begin{eqnarray}
	w(T)&=& w_0 \frac{T}{z_R}\,,\\
	\partial_x &\equiv& -  \frac{i T}{z_R} x\,, \\
	\partial_y &\equiv& -  \frac{i T}{z_R} y \,,
\end{eqnarray}
where
\textcolor{black}{
	\begin{eqnarray}
		T= 2 l a,
	\end{eqnarray}
}
For the case that we are interested in, $a$ is large. So we approximate
\eqref{GfactorWeakGravity} for large $a$, where the dominant term is given by $\cal O$:
\begin{eqnarray}
	{\cal O} &=&- \frac{19 i a^2}{2 z_R}(x^2-y^2) \,.
\end{eqnarray}
We observe that, for $w_0\ll \frac{z_R}{a}$, that holds true for a large value of $z_R$, the width of the package near the Sun remains much smaller than the Sun's radius.  We assume that $\sigma$ is not very sharp, so we ignore terms proportional to $x^-$. We assume that $a$ is large. So $\cal O$ gives the largest contribution.  We, therefore, keep only $\cal O$ and write: 
\begin{eqnarray}\label{SunPrediction}
	\psi_{\text{Bob}} &=&\left(1-i   \frac{9.5 m_\odot a^2}{z_R}(x^2-y^2)\right)  \psi_{\text{Alice}}|_{x^+=T}\\
	&=&\left(1-i   \frac{\textcolor{black}{28}~\text{km} \times a^2}{z_R}(x^2-y^2)\right)  \psi_{\text{Alice}}|_{x^+=T}\\
	&=&\left(1-  i \frac{1.12 \times 10^{9}~\text{meters}}{z_R}(x^2-y^2)\right)\psi_{\text{Alice}}|_{x^+=T} \,.
\end{eqnarray}
For $z_R<\textcolor{black}{28}~\text{km}\times a^2 $, the correction becomes larger than $1$ and we need to take into account higher $\epsilon$ terms. In order to keep the correction perturbative, we should choose either a sufficiently large value of $z_R$ or a null geodesic with smaller value of $a$. For $z_R=\textcolor{black}{2.8}~\text{km}\times a^2 = 1.34 \times 10^{9}~\text{m}$, the correction remains small and the wavepacket at Bob's position is,
\begin{eqnarray}
	\psi_{\text{Bob}}(x^\mu)&=& (1-  \textcolor{black}{0.10} i (x^2-y^2))  \psi_{\text{Alice}}(x^\mu)|_{x^+=T}\,.
\end{eqnarray}
The correction term $\epsilon \psi^{(1)}$, thus is,
\begin{equation}
	\epsilon\psi^{(1)}(x^\mu)= \textcolor{black}{0.10} i (x^2-y^2) \psi_{\text{Alice}}(x^\mu)|_{x^+=T}\,.
\end{equation}
We have depicted the amplitude and phase of the $\epsilon \psi^{(1)}$ in Fig.~4. We highlight that \eqref{SunPrediction} can be extended to the communication between Alice and Bob around the Earth as well,
\begin{eqnarray}\label{EarthPrediction}
	\psi_{\text{Bob}}(x^\mu)&=&\left(1-i   \frac{\textcolor{black}{9.5} m_\oplus a^2}{z_R}(x^2-y^2)\right)  \psi_{\text{Alice}}(x^\mu)|_{x^+=T}\\
	&=&\left(1-  i \frac{ \textcolor{black}{8.4}~\text{cm}\times a^2}{z_R}(x^2-y^2)\right)\psi_{\text{Alice}}(x^\mu)|_{x^+=T} \,.
\end{eqnarray}
We, however, see that for sufficiently large value of $z_R$ that satisfies all the constraints, the magnitude of the correction is far smaller than what we have predicted for the correction around the Sun.

\section*{\textbf{\large{Supplementary Note 7: On Gravitational Decoherence}}}
\begin{figure}
	\includegraphics[width=0.5\textwidth]{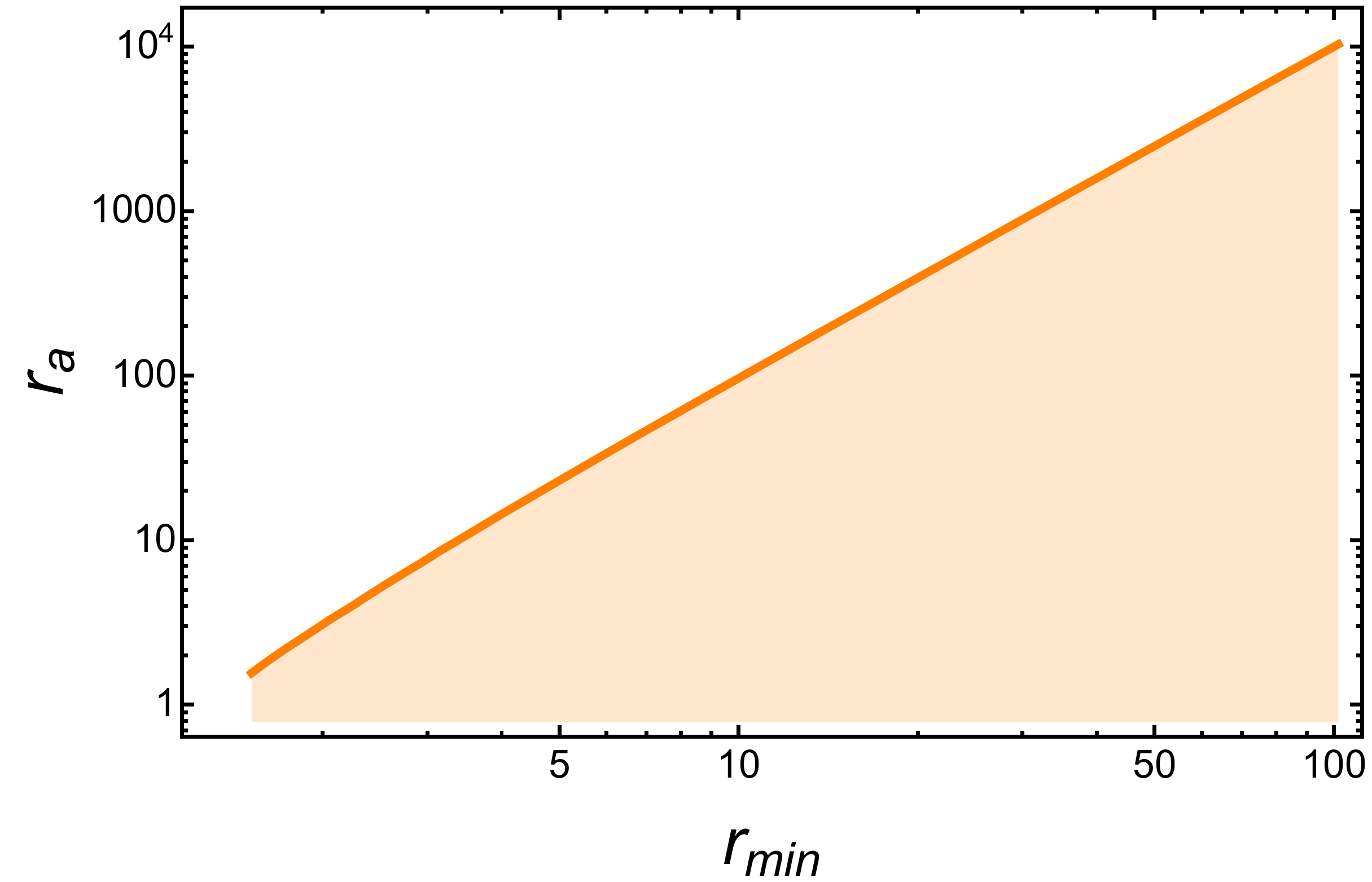}
	\caption{The shaded area shows the regime where $r_a$ and $r_{\text{min}}$ for which $|\tilde{\cal G}_{11}|$ remains smaller than 1. The unit of length is chosen equal to the Schwarzschild radius.}
	\label{fig:GT11NearEventHorizon}
\end{figure}
The consistency of the perturbation requires that all the $\epsilon$ terms in \eqref{OO}, \eqref{OQN} and \eqref{OQN}  remain smaller than 1. For a wavepacket with a frequency width of $\sigma$ and beam width of $W$, we can estimate the magnitude of each term by replacing $x$ with $W$, $\partial$ with $\frac{1}{W}$ and $x^-$ with $\frac{1}{\sigma}$. In particular, estimating the contribution of $\tilde{\cal G}_{11}$ in \eqref{GT11} to be smaller than 1, yields,
\begin{eqnarray}\label{Tmax}
	T  &\ll&\frac{2}{|{\cal G}_{11}|}\,,
\end{eqnarray}
where,
\begin{equation}
	T= 2\int_{r_{\text{min}}}^{r_a} \frac{dr}{\gamma{(r)}}\,,    
\end{equation}
is the time that the packet has felt the curved  spacetime geometry during its journey. ${\cal G}_{11}$ is a non-zero small parameter presented in \eqref{G11}. So \eqref{Tmax} identifies the maximum amount of time that  the wavepacket can retain its phase and coherence intact. For $r_\text{min} \ll r_a$ and large $l$, \eqref{G11_large_a} will prove that $|{\cal G}_{11}|=\frac{m}{l^2}$ and simplifies 
\eqref{Tmax} to,
\begin{eqnarray}\label{Tmax2}
	c T_{\text{Max}}  &\ll&\frac{l^2}{m}\,,
\end{eqnarray}
which shows that the curvature of the spacetime geometry ultimately changes the phase of the wavepacket moving on any null geodesic. For $T>T_{\text{Max}}$, Bob would need to take into account all the $\epsilon$ terms to recover the phase shift of the wavepacket that Alice has sent. This means that Bob must calculate the corrections up to an infinite derivatives of the  Riemann tensor. Taking into account all the derivatives of the Riemann tensor in a smooth geometry, is tantamount of knowing the exact value for the Riemann tensor in whole of the spacetime geometry, a piece of knowledge that is not  practically possible to gain. So it is tentative to argue that for $T>T_{\text{Max}}$, Bob practically does not have any chance to compute the effect of the correction due to the curvature of the spacetime geometry and recover what Alice has sent.  We refer to this phenomenon as practical gravitational decoherence.

For small values of $l$, the null geodesic goes very close to the event horizon of the black hole.  We have numerically performed the integration and calculated $\tilde{\cal G}_{11}$ for small values of $r_{min}$.  Fig. \ref{fig:GT11NearEventHorizon} shows the regime of $r_a$ and $r_{\text{min}}$ where $\tilde{\cal G}_{11}$ remains smaller than 1. We observe that when the wavepacket passes very close to the event horizon of the black hole, $\tilde{\cal G}_{11}$ becomes larger than one at around the black hole. This implies that a close encounter with the event horizon of the black hole completely changes phase of the wavepacket. The curvature of the spacetime geometry near the event horizon modifies the phase of the wavepacket to such a degree that the perturbation cannot be used to predict what far observers see. This could be interpreted as gravitational decoherence in the perturbative regime, and implies that Bob and Alice cannot easily communicate over a null geodesic that passes very close to the event horizon of the black hole, where the black hole adds lots of noise, and reducing the noise would be a computationally extensive or impossible task.

\section*{\textbf{\large{Supplementary Note 8:  On detecting the phase distortion near the Earth}}}

We assume that Alice sends a Gaussian wavepacket around $\omega=\omega_0$ with the width of $\sigma$ towards Bob. In other words, Alice at $r_a$ produces a wavepacket in the form of
\begin{equation}
	\label{Alice0}
	\Psi_{\text{Alice}}= f^{(0)}(x^+,x^1,x^2) e^{i\omega_0 x^-} e^{-\frac{(\sigma x^-)^2}{2}},
\end{equation}
and sends it towards Bob at $r_b$. The wavepacket moves along a radial null geodesic and reaches Bob. A radial null geodesic that represents an outgoing beam. It is described by $l=0$ in \eqref{RPaPbSimplified}. The only none-zero components of $R_{+\bar{a}+\bar{b}}$ for an outgoing beam, therefore, is,
\begin{eqnarray}
	R_{+-+-} = -\frac{1}{r^3}\,.
\end{eqnarray}
Equation~\eqref{rdot} implies that radial geodesic holds $\dot{r}=0$ which is solved to $x^+=r-r_a$ where Alice is located at $r_a$.  The only non-zero components of \eqref{eq21aux} contributing to the correction is,
\begin{equation}\label{GRadialSchwarzschild}
	{\cal G}_{--} = - \int\frac{d\tau}{r^3}= -\int_{r_a}^{r} \frac{dr}{\dot{r}r^3}=-\frac{1}{2}\left(\frac{1}{r_a^2}-\frac{1}{r^2}\right)\,.
\end{equation}
We can utilize \eqref{GRadialSchwarzschild} in  \eqref{OO}, \eqref{OQU} and \eqref{OQN}, and obtain,
\begin{subequations}
	\label{QORadial}
	\begin{eqnarray}
		{\cal O}^\omega &=& 0 \,, \\
		{\cal Q}^\omega_U &=& -\frac{i }{2\omega}\left(1-\frac{\left(\omega x^-\right)^2}{2}\right)\left(\frac{1}{r_a^2}-\frac{1}{r^2}\right)\,,\\
		{\cal Q}^\omega_N &=&-\frac{1}{2}\left(\frac{1}{r_a^2}-\frac{1}{r^2}\right) x^- \,. 
	\end{eqnarray}
\end{subequations}
The transverse structure of the wavepacket is not affected, as there is no derivative with respect to spatial transverse directions present in \eqref{QORadial}. This is due to high amount of symmetries present, the background is static and spherical, and the null geodesic respects the symmetries. 

In order to use the perturbative solution of the previous section, we notice that the curvature of the spacetime geometry during its journey has its largest value at Alice's position, where the Kretschmann invariant~\cite{Henry:1999rm} is,
\begin{equation}
	K= R_{\mu\nu\lambda\eta}R^{\mu\nu\lambda\eta}= \frac{12}{r^6_a}\,.
\end{equation}
We use the square root of the Kretschmann invariant as a measure to estimate the curvature of the spacetime geometry, i. e., $K \ll \omega_0^4$. We, therefore, impose,
\begin{subequations}
	\label{perturbativecondionsRadial}
	\begin{eqnarray}
		2\sqrt{3} &\ll& \omega_0^2 r_a^3  \,,\\
		2\sqrt{3} &\ll& \sigma^2 r_a^3  \,,
	\end{eqnarray}
\end{subequations}
then $\sigma \ll  \omega_0$ reduces \eqref{perturbativecondionsRadial} to 
\begin{subequations}
	\label{RCreducedSch}
	\begin{eqnarray}
		\label{RC2Sch}
		2\sqrt{3} &\ll& \sigma^2 r_a^3  \,,\\
		\label{RC3Sch}
		\sigma &\ll & \omega_0\,.
	\end{eqnarray}
\end{subequations}
Since $\sigma \ll \omega$,  ${\cal Q}^\omega_U$ and ${\cal Q}^\omega_N$ at Bob's position can be approximated to:
\begin{subequations}
	\begin{eqnarray}
		{\cal Q}^\omega_U &=& -\frac{i}{2\omega_0}\left(1-\frac{(\omega_0 x^-)^2}{2}\right)\left(\frac{1}{r_a^2}-\frac{1}{r_b^2}\right)\,,\\
		{\cal Q}^\omega_N &=&-\frac{x^-}{2}\left(\frac{1}{r_a^2}-\frac{1}{r_b^2}\right), 
	\end{eqnarray}
\end{subequations}
here, $\omega$ is approximated to $\omega_0$ in \eqref{QORadial}. Therefore, Bob receives the following wavepacket,
\begin{equation}
	\frac{\Psi_{\text{Bob}}}{f^{(0)}}= \Big(1+ \epsilon {\cal Q}^{\omega_0}_U+\epsilon {\cal Q}^{\omega_0}_{N}\Big)e^{i \omega_0 x^-} e^{-\frac{(\sigma x^-)^2 }{2}}+O(\epsilon^2)\,,
\end{equation}
where ${\cal Q}^\omega_N$ and ${\cal Q}^\omega_U$ alter the amplitude and phase of the wavepacket, respectively. The change in the amplitude of the wavepacket is given by,
\begin{equation}
	\delta {\cal A}=\epsilon Q^{\omega_0}_N e^{-\frac{(\sigma x^-)^2}{2}}+O(\epsilon^2)  = -\frac{\epsilon x^-}{2}\left(\frac{1}{r_a^2}-\frac{1}{r_b^2}\right)  e^{-\frac{(\sigma x^-)^2}{2}}+O(\epsilon^2) \,.
\end{equation}
The maximum change in the wavepacket amplitude occurs at $x^-=\pm\frac{1}{\sigma}$, and the maximum change yields
\begin{equation}
	|\delta {\cal A}_{\text{max.}}|=\frac{1}{2 \sigma}\left(\frac{1}{r_a^2}-\frac{1}{r_b^2}\right) \frac{1}{\sqrt{e}}\,.
\end{equation}
The consistency of the perturbation requires $\delta {\cal A}<1$, and thus, results in,
\begin{equation}\label{lb1}
	\frac{1}{4 e r_a}\left(1-\frac{r_a^2}{r_b^2}\right)^2  <\sigma^2 r_a^3\,.
\end{equation}
We notice that \eqref{lb1} is  already satisfied by  \eqref{perturbativecondionsRadial} since $1\ll r_a$. The imaginary part of the correction adds to the phase of the wavepackage, which is,
\begin{equation}
	\label{dPhiRadial}
	\delta {\chi}= \epsilon Q^{\omega_0}_U e^{-\frac{(\sigma x^-)^2}{2}}+O(\epsilon^2)  \approx \frac{i \omega_0 }{4\sigma^2 }\left(\frac{1}{r_a^2}-\frac{1}{r_b^2}\right) (\sigma x^-)^2 e^{-\frac{(\sigma x^-)^2}{2}}+O(\epsilon^2) \,.
\end{equation}
The maximum change in the wavepacket phase occurs at $x^-=\pm\frac{\sqrt{2}}{\sigma}$, which is,
\begin{equation}\label{dPhiRadialMaximum}
	\delta {\chi}_{\text{max.}}=\frac{\omega_0m_\oplus}{2  e  \sigma^2}\left(\frac{1}{r_a^2}-\frac{1}{r_b^2}\right) \,.
\end{equation}
where $m_\oplus$ is the Schwarzschild radius of Earth ($m_\oplus= \frac{2 G M_{\oplus}}{c^2}$) -- here, $c$ is recovered. Since the spacetime geometry is static, and considering that the equation we have solved are linear, there is no need to assume that the length of the wavepacket is small in the $x^-$ coordinate. We can extend the correction for small values of $\sigma$. This can be easily understood as considering segment of a large wavepacket in $x^-$ direction, solving the equation for each segment and adding the solutions for all segments. Let us assume that Alice is on the surface of the Earth with $r_a=6400$~km while Bob is in the International Space Station with $r_b=6800$~km. Equation \eqref{dPhiRadialMaximum} then \href{https://www.wolframalpha.com/input/?i=c/(4*nepper number*pi)*2G*(Earth Mass)/c^2*(1/(6400*km)^2-1/(6800km)^2)}{gives}:
\begin{equation}\label{dPhiISS}
	\delta {\chi}_{\text{max.}}=(2.2\times 10^{-10} \text{Hz})\frac{\nu_0}{\Delta\nu^2} \,,
\end{equation}
where the laser line-width $\Delta\nu=2\pi\sigma$, and $\nu_0=2\pi \omega_0$ are used. 

Note that the size of the wavepacket in time should not be larger than the distance between Alice and Bob, i.e., $400$~km, or equivalently
\begin{equation}\label{Delta749}
	\Delta \nu >749~\text{Hz}\,.
\end{equation}
In deriving (4) from (3), it is assumed that the metric inside the wavepacket can be approximated to a constant metric. Equation \eqref{RPaPbSimplified} gives the $\epsilon$ corrections to the metric, so they can be approximated to constant parameter inside the wavepacket. For radial null geodesic $L=0$, only $R_{+-+-}$ is non-vanishing.  For communication over radial null geodesic between $r_b$ and $r_a$ ($r_a<r_b$), the difference in the component of the metric $g_{\mu\nu}$ inside the wavepacket to be much than one gives
\begin{equation}
	\left|\frac{\Delta g_{\mu\nu}}{g_{\mu\nu}}\right| = 3\left|\frac{\Delta R_{+-+-}}{R_{+-+-}}\right| .
\end{equation}
Utilizing the \eqref{RPaPbSimplified}, and noting that the linewidth of $\Delta \nu$ corresponds to a wavepacket with size of  ${c}/{\Delta \nu}$ in time direction, then implies
\begin{equation}
	\left|\frac{3c}{r_a}\right|\ll \Delta \nu .
\end{equation}
Therefore, for $r_a=6400$~km, we get the lower bound of 
\begin{equation}\label{Delta149}
	\Delta \nu \gg 149~\text{Hz}
\end{equation}
There exists commercial portable continuous wave (cw) lasers with the ultra narrow linewidth of $1$~Hz for $\lambda=657$~nm (\href{https://www.menlosystems.com/products/ultrastable-lasers/ors-cubic/}{https://www.menlosystems.com/products/ultrastable-lasers/ors-cubic/}). However, in our calculation, we assume Alice generates a very good coherent Gaussian beam in the form of \eqref{Alice0} with linewidth of $1$~kHz. Notice that $\Delta \nu=1$~kHz satisfies both \eqref{Delta749} and \eqref{Delta149}. 

For example, Eq. \eqref{dPhiISS} for $\nu_0=456$~THz ($\lambda_0=657$ nm), and $\Delta\nu=1$~kHz gives a phase alteration of
\begin{equation}\label{dPhiISS2}
	\delta {\chi}_{\text{max.}}=0.10 \,,
\end{equation}
which is of a measurable magnitude.

Although measuring phase of a wavepacket with a high-precision is not a big challenge, in order to close this section we suggest the following strategy. We consider that Alice and Bob have the same equipment to produce wavepackets. Alice first sends a wavepacket in the form of \eqref{Alice0} toward Bob. Bob receives the wavepacket, measures its amplitude over time and identifies $\sigma$. Bob generates the wavepacket that Alice has sent, identifying as a local oscillator. We call this as the ``standard'' wavepacket. Alice prepares the second wavepacket in the form \eqref{Alice0}  and sends it to Bob. Alice sends the second wavepacket $\tau_A$ seconds after sending the first one. Bob in his frame would receive the second wavepacket $\tau_B$ seconds after getting the first package. $\tau_A$ and $\tau_B$ are not the same due to gravitational and relativistic redshift.  Bob, however, knows when Alice will send the second wavepacket and can compute when he would receive the wavepacket. Bob generates a standard wavepacket and measure its phase difference with the second wavepacket. The phase difference is a function of time, i.e., $\delta \chi_{\text{exp}}=\delta \chi_{\text{exp}}(t)$.

Equation \eqref{dPhiRadial} maps the quadratic term in $t$ to the distortion due to the curvature of the space-time geometry. So Bob fits $\chi_{\text{exp}}$ to the best quadratic polynomial in $t$:
\begin{equation}
	\delta \chi_{\text{exp}}= \left(a+\delta a\right) \left(\sigma t\right)^2 + \left(b + \delta b\right) \left(\sigma t\right) + \left(c + \delta c\right),
\end{equation}
and finds the numbers associated to $a,b,c$, and their errors $\delta a$, $\delta b$ and $\delta c$. Notice that $b$ encodes un-accounted Doppler or gravitational redshift while $a$ is due to \eqref{dPhiRadial}, and is given by:
\begin{equation}
	\label{aparameter}
	a = \frac{\omega_0 m_{\oplus}}{4\sigma^2 }\left(\frac{1}{r_a^2}-\frac{1}{r_b^2}\right),
\end{equation} 
The magnitude of \eqref{dPhiISS2} shows that $a$ can be measured. However, the atmosphere between ISS and the ground base can introduce an error of such a magnitude that would not allow us to measure $a$ between Earth and the ISS. The experiment, therefore, should be performed between two satellites which are in different latitudes. We can substitute $r_a$ and $r_b$ in \eqref{aparameter} with the latitudes of the two satellites, and \eqref{aparameter} describes the correction when they communicate over a radial null geodesic.  For example for radial communication between satellites with $r_a=6600$~km and $r_b=7000$~km, \eqref{dPhiRadialMaximum}  \href{https://www.wolframalpha.com/input/?i=c/(4*nepper number*pi)*2G*(Earth Mass)/c^2*(1/(6600*km)^2-1/(7000km)^2)}{is given} by:
\begin{equation}
	\delta {\chi}_{\text{max.}}=(1.98\times 10^{-10} \text{Hz})\frac{\nu_0}{\Delta\nu^2}\,,
\end{equation}
that $\nu_0=456$~THz, and $\Delta\nu=1$~kHz results 
\begin{equation}
	\delta {\chi}_{\text{max.}}=0.09\,,
\end{equation}
which is of a measurable magnitude and is free of atmosphere turbulence as there exists no air between the two satellites.

\end{document}